\documentclass[11pt]{article}

\usepackage{amssymb, fullpage, times}

\newcommand{\ket}[1]{| #1 \rangle}
\newcommand{\bra}[1]{\langle #1 |}
\newcommand{\ketbra}[1]{| #1 \rangle \langle #1 |}
\newcommand{\genketbra}[2]{| #1 \rangle \langle #2 |}
\newcommand{\braket}[2]{\langle #1 | #2 \rangle}

\newcommand{\norm}[1]{\|{ #1 }\|}
\newcommand{\trnorm}[1]{\|{ #1 }\|_{{\rm tr}}}

\newcommand{\ovlap}[2]{{\rm ovlap}(#1, #2)}
\newcommand{\rank}[1]{{\rm rank}(#1)}
\newcommand{\core}[1]{{\rm c}(#1)}
\newcommand{\corefamily}[1]{{\rm ncf}(#1)}

\newcommand{\irr}[0]{{\rm Irr}}
\newcommand{\linspan}[1]{{\rm span} \{#1\}}
\newcommand{\totvar}[1]{\|{ #1 }\|_1}

\newcommand{\QFT}[0]{{\rm QFT}}

\newcommand{\R}{{\mathbb{R}}}
\newcommand{\C}{{\mathbb{C}}}
\newcommand{\F}{{\mathbb{F}}}
\newcommand{\Z}{{\mathbb{Z}}}

\newcommand{\U}{{\mathbf{U}}}

\newcommand{\cHp}{{\cal H}_p}
\newcommand{\cB}{{\cal B}}

\newcommand{\onemat}{\leavevmode\hbox{\small1\kern-3.8pt\normalsize1}}

\newcommand{\ind}[0]{\uparrow}

\newcommand{\qed}{\hfill{$\rule{6pt}{6pt}$}} %Box at end of proof
\newenvironment{proof}{\noindent{\bf Proof}:}{\qed}

\newtheorem{lemma}{Lemma}
\newtheorem{proposition}{Proposition}
\newtheorem{theorem}{Theorem}
\newtheorem{corollary}{Corollary}

\newtheorem{fact}{Fact}
\newtheorem{definition}{Definition}

\begin{document}

\title{\Large \textbf{On the Power of Random Bases in Fourier
Sampling:\\ Hidden Subgroup Problem in the Heisenberg Group}}

\author{
Jaikumar Radhakrishnan
\thanks{School of Technology and Computer Science,
Tata Institute of Fundamental Research, Mumbai, India and
Toyota Technological Institute, Chicago, U.S.A.
Email: \texttt{jaikumar\symbol{64}tifr.res.in}}
\and
Martin R{\"o}tteler
\thanks{
NEC Laboratories America, Inc.,
Princeton, U.S.A.
Email: \texttt{mroetteler\symbol{64}nec-labs.com}}
\and
Pranab Sen
\thanks{Institute for Quantum Computing,
University of Waterloo, Canada.
Email: \texttt{p2sen\symbol{64}iqc.ca}}
}

%\date{\today}
\date{}

\maketitle

\begin{abstract}
The hidden subgroup problem (HSP) provides a unified framework to
study problems of group-theoretical nature in quantum computing such
as order finding and the discrete logarithm problem. 
While it is known that
Fourier sampling provides an efficient solution in the abelian case, 
not much is known 
for general non-abelian groups. Recently, some authors raised the
question as to whether post-processing the Fourier spectrum 
by measuring in a random orthonormal basis
helps for solving the HSP. Several
negative results on the shortcomings of this {\em random strong}
method are known.
In this paper however, we show that the random strong method 
can be quite
powerful under certain conditions on the group $G$. We
define a parameter $r(G)$ for a group $G$ and
show that $O((\log |G| / r(G))^2)$ iterations of the random strong
method give enough classical information to identify a hidden
subgroup in $G$. 
We illustrate the power of the random strong method via a
concrete example of the HSP over finite Heisenberg groups.
We show that $r(G) = \Omega(1)$ for
these groups; hence the
HSP can be solved using polynomially many random strong
Fourier samplings
followed by a possibly exponential classical post-processing without
further queries.
The quantum part of our algorithm consists of a polynomial computation
followed by measuring in a random orthonormal basis.
%We also show that earlier methods for non-abelian HSP fail for
%the Heisenberg group. 
This gives the first example of a group where random
representation bases do help in solving the HSP and for which no
explicit representation bases are known that solve the problem with
$(\log G)^{O(1)}$ Fourier samplings.
As an interesting by-product of our work,
we get an algorithm for solving the {\em state identification problem}
for a set of nearly orthogonal pure quantum states.
\end{abstract}

%%%%%%%%%%%%%%%%%%%%%%%%%%%%%%%%%%%%%%%%%%%%%%%%%%%%%%%%%%%%
%
% Section: Introduction
%
%%%%%%%%%%%%%%%%%%%%%%%%%%%%%%%%%%%%%%%%%%%%%%%%%%%%%%%%%%%%

\section{Introduction}

The hidden subgroup problem (HSP) is defined as follows: We are given
a function $f : G \rightarrow S$ from a group $G$ to a set $S$ with
the promise that there exists a subgroup $H \leq G$ such that $f$ is
constant on the left cosets of $H$ and takes distinct values on
distinct cosets.  In this paper, all groups and sets are finite and
all vector spaces are finite dimensional over $\C$.  The function $f$
is given via a black box, i.\,e., given $x \in G$ as input, the black
box outputs $f(x)$.  The task is to find a set of generators for $H$
while making as few queries to $f$ as possible. We would also like our
algorithm to be efficient in terms of total running time.  The abelian
HSP (i.\,e. $G$ is abelian) encompasses several interesting problems
such as finding the order of an element in a group and the discrete
logarithm problem. Factoring an integer $n$ can be reduced to order
finding in the group $\Z^*_n$, the multiplicative group of integers
modulo $n$ which are coprime to $n$.  The problems of graph
isomorphism and graph automorphism can be cast as hidden subgroup
problems over the non-abelian group $S_n$, the group of permutations
on $n$ symbols.

The classical query complexity of the HSP is $|G|^{\Omega(1)}$ which
is exponential in the input size $\log |G|$. This is true for many
families of groups including several families of abelian groups. The
biggest success of quantum algorithms so far has been a polynomial
time (both query complexity as well as total running time) solution
for the abelian HSP~\cite{Kitaev:95, BH:97,
ME:98}.  The heart of this solution is Fourier sampling with respect
to the abelian group $G$.

In sharp contrast to the abelian HSP, progress on the non-abelian
HSP (i.\,e. $G$ is non-abelian) has so far been quite limited. 
Ettinger, H{\o}yer and Knill~\cite{EHK:2004} prove that the 
quantum query complexity of the
non-abelian HSP is $O(\log |G|)$; however, their algorithm
takes $2^{O(\log^2 |G|)}$ quantum operations.
Ivanyos et al.~\cite{IMS:2003} and Friedl et al.~\cite{FIMSS:2003} 
apply abelian Fourier transform methods
to give polynomial quantum algorithms for the HSP for some special
classes of non-abelian groups. 
%\cite{EH:98} perform Fourier sampling
%over the abelian group $\Z_n \times \Z_2$ to solve the HSP
%over the dihedral group $\Z_n \rtimes \Z_2$, a non-abelian group.
%Their algorithm makes $O(\log n)$ queries and
%has a polynomial initial quantum part followed by an
%exponential classical post-processing without further queries. 
Given the success of Fourier
sampling in solving the abelian HSP, one can similarly ask
whether Fourier sampling over the non-abelian group $G$ helps
in solving the HSP over $G$. The Fourier transform over
a (in general, non-abelian) group $G$ gives
us a superposition over $(\rho, i, j)$ where $\rho$
is an irreducible unitary representation of $G$ and $i, j$ are the
row and column indices of the matrix $\rho$. The choice of basis
for $\rho$ gives us a degree of freedom in defining the Fourier
transform over $G$. This is in contrast to the abelian case, where all
representations are one-dimensional and hence only their names
$\rho$ matter.
The algorthim starts out with a tensor product of $t = O(\log |G|)$
superpositions over random cosets of the hidden subgroup $H$.
Exploiting the symmetries in these states, one can show that
(see e.g.~\cite{Kuperberg:2003, Ip:2003, MRS:2005}) the optimal 
measurement to recover $H$ consists of applying the
Fourier transform to each coset state, measuring the
names of the $t$ irreducible representations, followed by
a joint POVM on the column spaces of the resulting $t$ states. 
In {\em strong Fourier sampling}, one measures
%changes MR, 03/10/05
each of the $t$ column spaces using an orthonormal basis, i.\,e.,
one performs a tensor product of $t$ complete von Neumann measurements
instead of a joint POVM. In {\em weak Fourier sampling}, one measures
the names of the $t$ representations only.

Hallgren, Russell and Ta-Shma~\cite{HRT:2003} showed that 
polynomially many iterations of 
weak Fourier sampling give enough information to reconstruct normal
hidden subgroups. More generally, they show that the normal core
$c(H)$ of the hidden subgroup $H$ (i.\,e. the largest normal subgroup
of $G$ contained in $H$) can be reconstructed via the weak
method. Grigni, Schulman, Vazirani and Vazirani~\cite{GSVV:2004} and
Gavinsky~\cite{Gav:2004} extended the weak method to
find a hidden subgroup $H$ in $G$ if 
$[G : \kappa(G) H] = (\log |G|)^{O(1)}$. Here, $\kappa(G)$ is the
Baer subgroup of $G$ defined as
$\kappa(G) = \bigcap_{K: K \leq G} N(K)$, 
where $N(K)$ denotes the normaliser of $K$ in $G$.
The main shortcoming of the weak method is
that it gives exactly the same probability distribution if the hidden
subgroup is $H$ or a conjugate $g H g^{-1}$ of $H$. This leads us to
consider the {\em strong} method.
The amount of additional information about the hidden
subgroup $H$ that can be extracted by measuring the column space
in an orthonormal basis depends, in general,
on the particular basis. In a recent paper,
Moore, Russell and Schulman~\cite{MRS:2005} showed that for 
the symmetric group $S_n$,
for any choice of bases for the representations,
there are order two subgroups that require exponential number
of strong Fourier samplings in order to distinguish them from the
identity subgroup.
Grigni et al.~\cite{GSVV:2004} study the {\em random strong} 
method where a
random measurement basis is used for each representation $\rho$. 
They define a
group-theoretic parameter $\alpha$ depending on $G$ and $H$ and show
that if $\alpha$ is exponentially large, the additional advantage of
the random strong method over the weak method is exponentially
small. In particular, this is case when $G = S_n$ and $H \leq S_n$,
$|H| = 2^{O(n \log n)}$.  

\subsection{Our contributions}
In this paper, we analyse the power of the random strong method
and show, for the first time, that under certain (different) 
general conditions
on $G$ polynomially many iterations of the random strong method do
give enough classical information to identify $H$. 
We illustrate the power of the random strong method via a
concrete example of the HSP over finite Heisenberg groups $\cHp$
of order $p^3$, where $p \geq 3$ is a prime.
$\cHp$ is defined as the following set of upper triangular
matrices:
\begin{equation}
\label{eq:heisenberg}
\cHp := \left\{ \left( 
\begin{array}{ccc}
1 & x & z \\
0 & 1 & y \\
0 & 0 & 1
\end{array}
\right) : x, y, z \in \F_p \right\}.
\end{equation}
A convenient encoding for the elements of $\cHp$ is to write
$(x, y, z)$, where $x, y, z \in \F_p$ match the components in
equation~(\ref{eq:heisenberg}). 
The composition of two elements is then given by
\[
(x_1,y_1,z_1)\circ(x_2,y_2,z_2)=(x_1+x_2,y_1+y_2,z_1+z_2+x_1 y_2),
\]
and the inverse of an element is given by $(x,y,z)^{-1} = (-x,-y,xy -
z)$.  It is easy to see that the classical randomised query complexity
of the HSP on $\cHp$ is $\theta(p)$. The generic quantum algorithm of
Ettinger, H{\o}yer and Knill~\cite{EHK:2004} achieves $O(\log p)$ 
query complexity, but at the
expense of $p^{O(\log p)}$ quantum operations.  An algorithm with
$2^{\theta(\sqrt{\log p})}$ quantum operations can be obtained by
combining the ideas of \cite{Kuperberg:2003} and
\cite{FIMSS:2003}. However, the query complexity of this algorithm is
also $2^{\theta(\sqrt{\log p})}$. It seems non-trivial to design a
quantum algorithm with $(\log p)^{O(1)}$ query complexity and total
running time $p^{O(1)}$. In the following paragraphs, we indicate how
various existing methods for non-abelian HSP fail to achieve 
this goal. After
that, we show how the random strong method attains this goal,
illustrating the power of random bases in Fourier sampling.

It can be shown that $\cHp$ is a semidirect product of
the form $\Z_p \ltimes (\Z_p \times \Z_p)$, where the normal subgroup
is given by $N_\infty := \{(0,y, z): y, z \in \F_p\}$ and the 
complement by $A_{0,0} := \{(x, 0, 0): x \in \F_p\}$.  
The commutator subgroup of $\cHp$ is
given by $[\cHp, \cHp] = \{(0, 0, z): z \in \F_p\}$, 
which is also the centre $\zeta(\cHp)$.
The commutator subgroup is isomorphic to $\Z_p$; hence it
is abelian but not {\em smoothly abelian}
(an abelian group $G$ is said to be smoothly 
abelian~\cite{FIMSS:2003}
if it is the direct product of a subgroup of bounded exponent
and a subgroup of size $(\log |G|)^{O(1)}$).
The Baer subgroup turns out to be $\kappa(\cHp) = \zeta(\cHp)$. 
If $A \leq \cHp$, $|A| = p$, then $|\kappa(\cHp) A| \leq p^2$;
therefore for such an $A$,
$[G : \kappa(\cHp) A] \geq p$. In fact, we will see later
that there are $(p^2 + p + 1)$ order $p$ subgroups of $\cHp$.
Thus, the methods
of \cite{GSVV:2004,Gav:2004,IMS:2003,FIMSS:2003} are not 
applicable in order to
solve the HSP for $\cHp$ efficiently.
For more details about the Heisenberg group, see
Section~\ref{sec:heisenberg}.

The chief obstacle to finding hidden subgroups in $\cHp$ arises from
the order $p$ subgroups of $\cHp$ other than its centre. There are
$(p^2 + p)$ such order $p$ subgroups; we shall call them $A_{i,j}$, $i
\in \F_p \cup \{\infty\}$, $j \in \F_p$.  The {\em forgetful abelian
method} (i.\,e.  Fourier sampling over the abelian group $\Z_p \times
(\Z_p \times \Z_p)$ instead of the non-abelian group $\cHp \cong \Z_p
\ltimes (\Z_p \times \Z_p)$), weak Fourier sampling, strong Fourier
sampling in the natural representation basis of $\cHp$ (i.\,e. the
representation basis adapted to the distinguised subgroup tower $\{1\}
\lhd N_\infty \lhd \cHp$) as well as strong Fourier sampling in the
$\Z_p$-Fourier transform of the natural representation basis give
exponentially small information about the index $i$ of $A_{i,j}$. For
more details, see Section~\ref{subsec:failure}. For now, we give an
intuitive description of the main difficulty posed by these subgroups.
Suppose the hidden subgroup is $A_{i,j}$ for some $i \in \F_p \cup
\{\infty\}$, $j \in \F_p$.  With exponentially high probability,
Fourier sampling over $\cHp$ gives us a representation uniformly at
random from one of the $(p - 1)$ irreducible representations $\rho_k$
of degree $p$ for $k = 1, \ldots, p - 1$ of $\cHp$. Suppose one such
representation $\rho_k$ shows up.  The state essentially collapses to
a vector $\ket{\psi_{k,i,j}} \in \C^p$,
i.\,e., $(\cHp,A_{i,j})$ is a Gelfand pair for all $i$,$j$ (see also
\cite{MR:2005} for Gelfand pairs in the context of the HSP).
The vectors
$\ket{\psi_{k,i,j}}$ have the property that
\[
|\braket{\psi_{k,i,j}}{\psi_{k,i',j'}}| =
\left\{ \begin{array}{ccl}
\frac{1}{\sqrt{p}} & : & i \not= i', \; \mbox{for all}\; j,
j', \\[1ex]
\delta_{j, j'} & : & i = i',
\end{array}
\right.
\]
i.\,e., they form a set of $(p + 1)$ mutually unbiased 
bases~\cite{WF:89} of $\C^p$.
The main difficulty is that it is not clear a priori
that there is any orthonormal basis that can pairwise distinguish
between these $(p^2 + p)$ vectors with inverse polynomial probability.
Note that the so-called
{\em hidden conjugate problem}~\cite{MRRS:2004} is easy to solve 
information-theoretically
for $\cHp$; the conjugacy classes of the order
$p$ subgroups are defined by $i$ and the above property says that
$\{\ket{\psi_{k,i,j}}\}_j$ is an orthonormal basis of $\C^p$, so
given the conjugacy class $i$ one can measure in this orthonormal
basis to determine the actual hidden subgroup $A_{i,j}$. 
In view of this, the main challenge in solving the HSP for 
$\cHp$ is to identify the conjugacy class $i$.

In this paper however, we show that a 
random representation basis for $\rho_k$ does in fact
pairwise distinguish between 
$\ket{\psi_{k,i,j}}$ with constant probability.
In fact, we refine the method of random measurement bases to
distinguish between families of nearly orthogonal subspaces.
We combine the geometric ideas of random measurement bases together
% changes PS, 03/10/05
with representation-theoretic techniques to 
obtain a parameter $r(G; H_1, H_2)$ of a group $G$ and subgroups
$H_1, H_2 \leq G$. We show that $r(G; H_1, H_2)$ is a lower
bound on the total variation distance between the distributions
on pairs $(\rho, j)$ of representation names and column indices
% changes PS, 03/10/05
obtained by the random strong method for 
candidate hidden subgroups $H_1, H_2$. The parameter $r(G; H_1, H_2)$ 
is defined 
in terms of the ranks and overlaps of the projectors
% changes PS, 03/10/05
obtained by averaging representations $\rho$ over 
$H_1, H_2$. 
% changes PS, 03/10/05
Define $r(G) := \min_{H_1, H_2} r(G; H_1, H_2)$, where
$H_1, H_2$ range over all pairs of subgroups of $G$.
We show that $O\left(\frac{\log s(G)}{r^2(G)}\right)$ 
iterations of the random strong method give sufficient classical
information to identify the hidden subgroup $H$, where $s(G)$
denotes the number of distinct subgroups of $G$. Note that
$s(G) \leq 2^{\log^2 |G|}$ for any group $G$.

We will see later in Section~\ref{sec:heisenberg} that $s(\cHp) =
O(p^2)$. In Section~\ref{sec:power}, we show that $r(\cHp) =
\Omega(1)$, implying that $O(\log p)$ iterations of the random strong
method give sufficient information to extract the hidden subgroup in
$\cHp$.  This gives us an algorithm solving the HSP over $\cHp$ with
$O(\log p)$ query complexity, $O(\log^3 p)$ quantum operations for
implementing the non-abelian Fourier transforms (see 
Section~\ref{subsec:qft}), $\tilde{O}(p^2)$ quantum operations 
to measure in a
random basis, and $\tilde{O}(p^4)$ classical post-processing
operations.  This gives the first example of a group where random
representation bases do help in solving the HSP and for which no
explicit representation bases are known that solve the problem with
$(\log p)^{O(1)}$ Fourier samplings.

As an interesting by-product of our work, we get an algorithm for
solving the following {\em quantum state identification} problem:
Consider a set of pure quantum states $\{\ket{\psi_1}, \ldots,
\ket{\psi_m}\} \in \C^n$ with the property that
$|\braket{\psi_i}{\psi_j}| \leq \delta$ for all $i \neq j$, where
$\delta$ is a sufficiently small constant (and typically $m \gg
n$). We are given $t$ independent copies of $\ket{\psi_i}$. The task
is to identify the index $i$. We show that $t = O(\log m)$ independent
random complete von Neumann measurements in $\C^n$ suffice to identify
$i$ with high probability.

%changes PS, 03/10/05
\subsection{Relation to other work}
Moore, Rockmore, Russell and Schulman~\cite{MRRS:2004}
use non-abelian strong Fourier sampling to give an efficient algorithm
for the HSP over the $q$-hedral group $\Z_q \ltimes \Z_p$ when
$p, q$ are prime, $q \mid (p - 1)$ and 
$(p - 1) / q = (\log p)^{O(1)}$. 
Our techniques show that for $p, q$ prime, $q \mid (p - 1)$,
$q = \Omega(\sqrt{p})$,
$r(\Z_q \ltimes \Z_p) = \Omega(1)$,
which proves that polynomially many random strong Fourier samplings
suffice to find an arbitrary hidden subroup of $\Z_q \ltimes \Z_p$
in this case.
For prime $p$, $q \mid (p - 1)$, $q = \Omega(p^{3/4})$,
subgroups $H_1, H_2$ conjugate to $\Z_q \leq \Z_{p-1}$, our
techniques show that
$r(\Z_{p-1} \ltimes \Z_p; H_1, H_2) = 
 \Omega\left(\sqrt{\frac{q}{p}}\right)$.
Moore et al.~\cite{MRRS:2004} prove a nearly matching upper bound
of $r(\Z_{p-1} \ltimes \Z_p; H_1, H_2) =
O\left(\sqrt{\frac{q}{p}\log{p}} \right)$. Thus, a
polynomial amount of
random strong Fourier sampling can solve the hidden conjugate
problem for subgroup $\Z_q \leq \Z_{p-1}$ of the affine group
$\Z_{p-1} \ltimes \Z_p$ if and only if $p/q = (\log p)^{O(1)}$.

In this paper, we confine ourselves to random strong Fourier sampling.
Our quantum operations always factor into a tensor product over the
coset states obtained by querying the function oracle. This 
distinguishes the Heisenberg
group from the symmetric group for which 
Moore, Russell and Schulman~\cite{MRS:2005} show that tensor 
product Fourier sampling
is not sufficient to solve the HSP. The quantum part of our algorithm
consists of a polynomial computation
followed by measuring in a random orthonormal basis. In
fact, if a suitable kind of pseudo-random unitary transformation 
can be generated and implemented efficiently,
then the quantum part of the algorithm
can be made fully polynomial. Various notions of pseudo-random unitary
transformations have been studied 
(see e.g.~\cite{EWSLC:2003, Emerson:2004}), 
but it has to be investigated whether they are
sufficient for our purposes.

%%%%%%%%%%%%%%%%%%%%%%%%%%%%%%%%%%%%%%%%%%%%%%%%%%%%%%%%%%%%
%
% Section: Heisenberg groups over $\F_p$
%
%%%%%%%%%%%%%%%%%%%%%%%%%%%%%%%%%%%%%%%%%%%%%%%%%%%%%%%%%%%%

\section{Heisenberg groups over $\F_p$}
\label{sec:heisenberg}
The groups $\cHp$, where $p\geq 3$ is prime, are discrete versions of
the continuous Heisenberg groups studied in physics in the context of
conjugate observables.  Abstractly, $\cHp$ is isomorphic to the
following group given in terms of generators and relations: 
$\cHp \cong \langle x, y, z : x^p = y^p = z^p = 1, xy = zyx, xz = zx,
                              zy = yz \rangle$.

\subsection{The subgroup lattice}
Since the order of $\cHp$ is $p^3$ we can expect to find
subgroups of order $p$ and $p^2$ besides the trivial subgroup
$\{1\}$ and $\cHp$.
The centre of $\cHp$ is given by 
\[
\zeta(\cHp) = \langle (0, 0, 1) \rangle 
            = \{(0, 0, z): z \in \F_p\}. 
\]
Note that $|\zeta(\cHp)| = p$.
There are $p+1$ subgroups $N_i$ of
order $p^2$, where $i \in \F_p \cup \{\infty\}$. They are given
by 
\[
N_i := \langle (1, i, 0), (0,0,1) \rangle
     = \{(x, xi, z): x, z \in \F_p \}, 
~~~ \forall i \in \F_p.
\]
The group $N_\infty$ is given by $ N_\infty := \langle (0, 1, 0),
(0,0,1) = \{(0, y, z): y, z \in \F_p\}$; 
$N_\infty \cong \Z_p \times \Z_p$. 
It is easy to see that for
all $i \in \F_p \cup \{\infty\}$, $\zeta(\cHp) \lhd N_i$. 
Furthermore the $N_i$ are normal subgroups, $N_i \lhd \cHp$ 
and $N_i \cong \Z_p \times \Z_p$.
For each $i \in \F_p \cup \{\infty\}$, we have that $N_i$ contains $p$
subgroups $A_{i,j}$ for $j \in \F_p$. The subgroups $A_{i,j}$ satisfy
$|A_{i,j}| = p$, whence $A_{i,j} \cong \Z_p$.  For $i, j \in \F_p$ we
have the following explicit desciption of the elements of $A_{i,j}$:
\[
A_{i,j} := \langle (1, i, j) \rangle
         = \{(\mu, \mu i, {\mu \choose 2} i + \mu j): \mu \in \F_p\}.
\] 
For $i = \infty, j \in \F_p$ we obtain $ A_{\infty, j} := \langle (0,
1, j) \rangle = \{(0, \mu, \mu j): \mu \in \F_p\}$. It is easy to see
that $A_{i,j} \not \leq N_{i'}$ if $i \neq i'$, and 
the normaliser is given by $N_{{\cal H}_p}(A_{i,j}) = N_i$.  
%For each
%$i \in \F_p \cup \{\infty\}$ and $j \in \F_p$, it is possible 
%to choose
%transversals for $A_{i,j}$ in $N_i$ in a uniform way so that they form
%a subgroup of $\cHp$\,: for example, we can choose as transversals the
%elements of the centre $\zeta(\cHp) = \{(0,0,z): z \in \F_p\}$.  For a
%coset $t A_{i,j}$, where $t = (0,0,\tau)$, $\tau \in \F_p$ we get
%\[
%t A_{i,j} = \{(\mu, \mu i, {\mu \choose 2}  i + \mu j + \tau) : 
%               \mu \in \F_p\}.
%\]
%For each $i \in \F_p$, we can choose the elements of the subgroup
%$A_{\infty,0} = \{(0,y,0): y \in \F_p\}$ as transversals of $N_i$ in
%$\cHp$.  We can choose the elements of the subgroup $A_{0,0} =
%\{(x,0,0): x \in \F_p\}$ as transversals of $N_\infty$ in $\cHp$. 
The above groups form a complete list of distinct subgroups of $\cHp$.
The following table summarizes the subgroup structure of $\cHp$.
\renewcommand{\arraystretch}{1.3}
\begin{center}
\begin{tabular}{|c|c|c|c|}
\hline
Size & Subgroup & Number & Containment \\
\hline
$p^3$ & $\cHp$ & $1$ & \\
\hline
$p^2$ & $N_i, i \in \F_p \cup \{\infty\}$ & $p+1$ & $N_i \lhd \cHp$\\
\hline
$p$   & $\zeta(\cHp)$, 
        $A_{i,j}, i \in \F_p \cup \{\infty\}, j \in \F_p$ 
      & $p^2+p+1$ & $A_{i,j} \lhd N_i$, 
                    $\zeta(\cHp) \lhd N_k, 
                     \forall k \in \F_p \cup \{\infty\}$ \\
\hline
$1$   & $\{1\}$ & $1$ & $\{1\} \lhd \zeta(\cHp)$,
                        $\{1\} \lhd A_{i,j}$ \\
\hline
\end{tabular}
\end{center}
\renewcommand{\arraystretch}{1}

For $i, i' \in \F_p \cup \{\infty\}$ where $i \neq i'$ we have that 
$N_i \cap N_{i'} = \zeta(\cHp)$. This shows that
$\kappa(\cHp) = \bigcap_{K: K \leq \cHp} N(K) = \zeta(\cHp)$. 
Also, it is easy to check that the commutator subgroup is
given by $[\cHp, \cHp] = \zeta(\cHp)$.

\subsection{The irreducible representations of $\cHp$}
Since we want to perform Fourier analysis on the groups $\cHp$
we have to determine the irreducible representations of
$\cHp$. The reader not familiar with the standard notations of
representation theory is referred 
%to Appendix \ref{ap:repTheory} or 
to standard references like \cite{CR:62} or \cite{Serre:77}. Observe
that 
$\cHp   =   A_{0,0} \ltimes N_\infty  
      \cong \Z_p \ltimes (\Z_p \times \Z_p)$. 
This semidirect product structure can be used
to construct the irreducible representations of $\cHp$.  First,
there are $p^2$ one-dimensional representations 
$\chi_{a,b}$ for $a,b \in \F_p$ which come from the 
factor group $\cHp / \zeta(\cHp) \cong \Z_p^2$. In the following,
let $\omega$ denote
a fixed $p$th root of unity in the complex numbers. Then the
one-dimensional irreducible representations of $\cHp$ are given by
\[
\chi_{a,b}((x,y,z)) := \omega^{ax+by} ~~~ a,b \in \F_p.
\]

Let $\F_p^\ast$ denote the group of non-zero elements of $\F_p$ under
multiplication.  There are $p-1$ irreducible representations $\rho_k$,
$k \in \F_p^\ast$ of degree $p$. They are obtained in the following
way: Take a nontrivial character of the centre $\zeta(\cHp)$, extend
it to the abelian group $N_\infty$, and induce it to
$\cHp$. Explicitly, we obtain the following representations: For each
$k \in \F_p^\ast$, we have a nontrivial character $\phi_k$ of
$\zeta(\cHp)$ given by $\phi_k((0,0,z)) := \omega^{kz}$. Since
$\zeta(\cHp) \lhd N_\infty$ and $N_\infty$ is abelian, we can extend
$\phi_k$ to a character $\overline{\phi}_k$ of $N_\infty$ by simply
defining $\overline{\phi}_k((0,y,0)) := 1$. We choose the elements of
$A_{0,0}$ as transversals for $N_\infty$ in $\cHp$. Then $\rho_k$ is
defined to be the induction $\rho_k := \overline{\phi}_k
\ind_{A_{0,0}} \cHp$.
%(see Appendix~\ref{ap:repTheory} for the 
%definition of induced representations) 
On the generators of $\cHp$,
we find that $\rho_k$ takes the following values: $\rho_k((1,0,0)) =
\sum_{a \in \F_p} \genketbra{a}{a+1}$, $\rho_k((0,1,0)) = \sum_{a \in
\F_p} \omega^{k a} \ketbra{a}$ and $\rho_k((0,0,1)) = \omega^{k}
\onemat_p$, where $\onemat_p$ denotes the identity operator in $\C^p$.
Since $(x,y,z)=(0,0,z)(0,y,0)(x,0,0)$ for all $x,y,z \in \F_p$, we
obtain that 
\[
\rho_k((x,y,z)) = \omega^{kz} \sum_{a \in \F_p} 
                  \omega^{k y a} \genketbra{a}{a+x}.
\]
It can be readily checked that the $\chi_{a,b}$, for $a, b \in \F_p$
and $\rho_k$, for $k \in \F_p^\ast$ form
a complete set of inequivalent irreducible representations of
$\cHp$.
%We now determine the Fourier coefficients of a coset state
%of a hidden subgroup $A_{i,j}$
%Denoting the coset representative $t = (0,0,c)$ we find 
%that the value
%of $\rho_k(t A_{i,j})$ is given by
%\begin{equation}\label{megaSum}
%\rho_k(t A_{i,j}) = \sum_{\mu = 0}^{p-1} \rho_k((\mu, \mu i, {\mu
%  \choose 2} i + \mu j + c)).
%\end{equation}

\subsection{Ranks and overlaps of various projectors}
Define $P_{k;i,j} := \frac{1}{p} \sum_{a \in A_{i,j}} \rho_k(a)$. 
It is easy to see that $P_{k;i,j}$ is an orthogonal projection
% changes MR, 03/10/05
operator.
In order to calculate the parameter $r(\cHp)$ (see
Section~\ref{sec:power} for the details of the calculation) we have to
compute the ranks of $P_{k;i,j}$ and pairwise {\em overlaps}
$\norm{P_{k;i,j} P_{k;i',j'}}$ (the reason for the nomenclature of
{\em overlap} will be made clear in Section~\ref{subsec:nearlyortho}).
%First, recall the following standard
%fact about $\cHp$.
%\begin{fact}
%The group $\cHp$ is a group of central type, i.\,e., for $x \in
%\cHp \setminus \zeta(\cHp)$ we obtain that $\chi(x) = 0$
%for all $\chi\in {\rm Irr}(\cHp)$ of ${\rm deg}(\chi) = p$
%(i.\,e., $\chi$ vanishes off the centre for irreducible
%representations of
%degree $p$).
%\end{fact}
%We can use this to show that $\rho_k(A_{i,j})$ has rank one. Let
%$\chi_k$ denote the character corresponding to $\rho_k$. Since
%$A_{i,j} \cap \zeta(\cHp) = \{ 1_{\cHp} \}$ we obtain that
%\[ 
%\rank{\rho_k(A_{i,j})} = \frac{1}{|A_{i,j}|} \langle \chi_k 
%\downarrow
%  A_{i,j} | 1 \rangle = \frac{1}{p} \sum_{x \in A_{i,j}} \chi_k(x) =
%  \frac{1}{p} \chi_k(1_{\cHp}) = 1.
%\]
%First, note that 
%because of the identity
%\[ 
%\left( 
%\begin{array}{ccc}
%1 & x & z \\
%0 & 1 & y \\
%0 & 0 & 1
%\end{array}
%\right)
%=
%\left( 
%\begin{array}{ccc}
%1 & 0 & z \\
%0 & 1 & 0 \\
%0 & 0 & 1
%\end{array}
%\right)
%\left( 
%\begin{array}{ccc}
%1 & 0 & 0 \\
%0 & 1 & y \\
%0 & 0 & 1
%\end{array}
%\right)
%\left( 
%\begin{array}{ccc}
%1 & x & 0 \\
%0 & 1 & 0 \\
%0 & 0 & 1
%\end{array}
%\right)
%\] 
For $i, j \in \F_p$, we obtain 
by a straightforward computation that
$P_{k;i,j} = 
 \frac{1}{p} \sum_{\mu,\nu \in \F_p}
 \omega_p^{k\left({\mu \choose 2}i + \mu j - {\nu \choose 2} i - 
                   \nu j\right)} 
 \ket{\nu}\bra{\mu}$.  
Hence,  
$P_{k;i,j} = \ketbra{\psi_{k;i,j}}$, where 
\[
\ket{\psi_{k,i,j}} =
\frac{1}{\sqrt{p}} \sum_{\mu \in \F_p} 
\omega^{-k \left({\mu \choose 2}i + \mu j\right)} \ket{\mu},
~~~
i, j \in \F_p, k \in \F_p^\ast. 
\]
In the case $i = \infty$, $j \in \F_p$, we get $P_{k;\infty,j} =
\ketbra{\psi_{k;\infty,j}}$, where $ \ket{\psi_{k;\infty,j}} =
\ket{-j} ~~~ j \in \F_p, k \in \F_p^\ast$.  Thus for all $k \in
\F_p^\ast$, $i \in \F_p \cup \{\infty\}$, $j \in \F_p$,
$\rank{P_{k;i,j}} = 1$ and $P_{k;i,j}$ is an orthogonal projection
onto $\ket{\psi_{k;i,j}}$.  For $j, j' \in \F_p$, we get
$\norm{P_{k;\infty,j} P_{k;\infty,j'}} = \delta_{j,j'}$.  For $i, i',
j' \in \F_p$, we get $\norm{P_{k;i,j} P_{k;\infty,j'}} =
\frac{1}{\sqrt{p}}$.  For $i, i', j, j' \in \F_p$, we get
\[
\norm{P_{k;i,j} P_{k;i',j'}} = 
|\braket{\psi_{k;i,j}}{\psi_{k;i',j'}}| =
\frac{1}{p} \sum_{\mu \in \F_p}
\omega^{k \left({\mu \choose 2} (i - i') + \mu (j - j')\right)}.
\]
To evaluate the last term above, we need the following fact
about quadratic Weil sums in $\F_p$.
\begin{fact}[\mbox{\cite[Theorem 5.37]{LN:94}}]
\label{fact:weil}
Let $h(X) \in \F_p[X]$ be a
degree two polynomial. Then, 
$\displaystyle
\left|\sum_{x \in \F_p} \omega^{h(x)}\right| = \sqrt{p}$.
\end{fact}
By Fact~\ref{fact:weil}, if $i \neq i'$,
$|\braket{\psi_{k;i,j}}{\psi_{k;i',j'}}| = \frac{1}{\sqrt{p}}$ 
irrespective of $j$ and $j'$. 
If $i = i'$, it is easy to see that
$|\braket{\psi_{k;i,j}}{\psi_{k;i',j'}}| = \delta_{j,j'}$.
To summarise, we have shown the following result:
\begin{lemma}
\label{lem:ovlapHeisen}
Suppose $p$ is an odd prime.  Let $i, i' \in \F_p \cup \{\infty\}$,
$j, j' \in \F_p$ and $A_{i,j}, A_{i',j'}$ be two order $p$ subgroups
of $\cHp$ other than the centre $\zeta(\cHp)$.  Let $\rho_k$, where $k
\in \F_p^\ast$, be an irreducible representation of $\cHp$ of degree
$p$. Let $P_{k;i,j}$ be defined by $P_{k;i,j} := \frac{1}{p} \sum_{a
\in A_{i,j}} \rho_k(a)$ and let $P_{k;i',j'}$ be defined
similarly. Then $P_{k;i,j}$, $P_{k;i',j'}$ are rank one orthogonal
projections, and their overlap is given by
\[
\norm{P_{k;i,j} P_{k;i',j'}} =
\left\{
\begin{array}{ccl}
\frac{1}{\sqrt{p}} & : & i \not= i', \; \mbox{for all}\; j,j',\\[1ex]
\delta_{j, j'}     & : & i = i'.
\end{array}
\right.
\]
\end{lemma}
Thus, for any $k \in \F_p^\ast$, the vectors $\ket{\psi_{k;i',j'}}$
form a set of $(p + 1)$ mutually unbiased bases for $\C^p$.

\subsection{Failure of existing methods to solve the HSP over $\cHp$}
\label{subsec:failure}
A straightforward classical randomised algorithm for the HSP over
$\cHp$ is as follows: Query $f: \cHp \rightarrow S$ at $O(p)$
random elements of $\cHp$.  If we do not find $a_1, a_2 \in \cHp$,
$a_1 \neq a_2$ such that $f(a_1) = f(a_2)$, we declare $\{1\}$ to be
the HSP of $f$. Suppose we do find such a pair $a_1, a_2$. Then there
is a unique order $p$ subgroup $A$ of $\cHp$ such that $a_1^{-1} a_2
\in A$.  $f$ can now be thought of as a function on $\cHp / A$. Query
$f$ at $O(\sqrt{p})$ random elements of $\cHp / A$.  If we do not find
$b_1, b_2 \in \cHp / A$, $b_1 \neq b_2$ such that $f(b_1) = f(b_2)$,
we declare $A$ to be the HSP of $f$. Suppose we do find such a pair
$b_1, b_2$. Let $B = \langle A, b_1^{-1} b_2 \rangle$.  If $|B| =
p^3$, declare the HSP to be $\cHp$.  If $|B| = p^2$, query $f$ at an
element $c \in \cHp$, $c \not \in B$.  If $f(c) = f(B)$, declare the
HSP to be $\cHp$, else declare the HSP to be $B$. The correctness of
the algorithm follows from the subgroup structure of $\cHp$ and the
birthday paradox. A matching lower bound of $\Omega(p)$ for classical
randomised algorithms can be proved using the subgroup structure of
$\cHp$ and Yao's minimax principle.

Suppose the HSP is $A_{i,j}$, for some $i \in \F_p^\ast$, $j \in
\F_p$.  It can be shown (see Section~\ref{subsec:hspintro} for
details) that Fourier sampling gives a $p$-dimensional representation
with probability $1 - \frac{1}{p}$, and each $p$-dimensional
representation has equal probability to show up.  Suppose one such
representation $\rho_k$, $k \in \F_p^\ast$ shows up.  The natural
representation basis $\ket{a}$, $a \in \F_p$ is the basis
$\ket{\psi_{k;\infty,j}}$, where $j \in \F_p$.  The $\Z_p$-Fourier
transform of the natural representation basis is the basis
$\ket{\psi_{k;0,j}}$, where $j \in \F_p$. By
Lemma~\ref{lem:ovlapHeisen}, the probability distribution obtained by
measuring the columns of $\rho_k$ in the natural representation basis
or in the $\Z_p$-Fourier transform of the natural representation basis
is the uniform distribution.  This shows that weak Fourier sampling,
strong Fourier sampling in the natural representation basis of $\cHp$
as well as strong Fourier sampling in the $\Z_p$-Fourier transform of
the natural representation basis give exponentially small information
about the index $i$ of $A_{i,j}$.

%In \cite{EH:98} an algorithm for the hidden subgroup problem in
%dihedral groups was given which is polynomial in the number of 
%quantum operations, however, exponential in the classical
%post-processing. The quantum part consists of a remarkably simple
%transformation. Despite the fact that the dihedral group is a
%nonabelian semidirect product of the form $\Z_2 \ltimes \Z_N$, the
%authors show that the classical data obtained from measuring in a
%Fourier basis for the abelian group $\Z_2 \times \Z_N$ is enough to
%reconstruct a hidden reflection with high probability from 
%$O(\log N)$
%samples. This method has been referred to as the ``forgetful'' method
%since it ignores the structure of the underlying group. In
%\cite{MRRS:2004} it has been shown that for the affine groups
%$\AGL(2,\F_p)$ which are isomorphic to 
%$\Z_{p-1} \ltimes \Z_p$ a Fourier
%transform for the group $\Z_{p-1} \times \Z_p$ does not help even to
%solve the easier problem of distinguishing between conjugate
%subgroups. 
%
Recall that $\cHp = A_{0,0} \ltimes N_\infty \cong \Z_p \ltimes (\Z_p
\times \Z_p)$.  Suppose we try to perform Fourier sampling over the
abelian group $A_{0,0} \times N_\infty \cong \Z_p \times \Z_p \times
\Z_p$ (the {\em forgetful abelian method}) instead of the non-abelian
group $\cHp$. Let $F$ denote the Fourier transform over $\Z_p \times
\Z_p \times \Z_p$, i.\,e.,
\[
F = p^{-3/2} \sum_{a,b,c,x,y,z \in \F_p} \omega^{ax+by+cz}
              \ket{a,b,c}\bra{x,y,z}.
\]
For abelian groups $G$, the probability distributions obtained by
Fourier sampling over $G$ are independent of the actual coset of the
hidden subgroup that arises on measuring the function value; however,
they depend of course on the hidden subgroup. But since now we are
doing abelian Fourier sampling over a non-abelian group, we have to
consider the effect of applying $F$ to a coset of $t A_{i,j}$, where
$t = (0,0,\tau)$ and $\tau \in \F_p$.  Note that $t A_{i,j} =
\{(\mu, \mu i, {\mu \choose 2} i + \mu j + \tau) : \mu \in \F_p\}$.
We obtain
\[
F \ket{t A_{i,j}} =
%\sum_{a,b,c, x,y,z} \omega_p^{ax+by+cz}
%\ket{a,b,c}\bra{x,y,z} \sum_{\mu=0}^{p-1} |{\mu, \mu i, {\mu
%\choose 2} i + \mu j + \tau}\rangle \\
%& = & \frac{1}{p^2} \sum_{a,b,c \in \F_p} \left(\sum_{\mu \in \F_p} 
%      \omega^{a\mu + b \mu i + c\left({\mu \choose 2} i + \mu j + 
%              \tau\right)} 
%      \right) \ket{a,b,c} \\
%& = & 
\frac{1}{p^2} \sum_{a,b,c \in \F_p} \omega^{c \tau} 
\left(\sum_{\mu \in \F_p}
\omega^{(a+bi+cj-\frac{ci}{2})\mu + ci \mu^2} 
\right) \ket{a,b,c}.
\]
Hence, the probability of observing a particular triple $(a,b,c)$ is
$p^{-4} \big|\sum_{\mu \in \F_p} \omega^{(a+bi+cj-(ci)/2)\mu + ci
\mu^2} \big|^2$.  If $c \neq 0$, this is a quadratic Weil sum and we
can use Fact~\ref{fact:weil} to conclude that the probability of
observing $(a,b,c)$ is given by $p^{-3}$, independent of $i,j$.  The
probability of observing $(a,b,c)$, $c \neq 0$ is $1 - \frac{1}{p}$.
If $c = 0$, only terms of the form $(-bi, b, 0)$ show up.  These terms
do give information about $i$; however, the probability of observing
such a term is $\frac{1}{p}$.  Thus, the forgetful abelian method
gives exponentially small information about $i$.

\subsection{Efficient quantum circuits for the Fourier Transform on
$\cHp$}
\label{subsec:qft}
The fact that any $\QFT$ for any finite group is a unitary matrix
(when properly normalized) makes this class of transformations an
important source of transformations a quantum computer can carry
out. The problem of finding efficient implementations of $\QFT$s in
terms of quantum circuits was studied previously, see
\cite{Hoyer:97,Beals:97,PRB:99,HRT:2003,MRR:2004}. 
From \cite[Theorem 2]{MRR:2004} it follows that for any prime $p$ 
the $\QFT$ for the
Heisenberg group $\cHp$ can by computed in ${\rm polylog}(p)$
operations. In the following we give an explicit description of an
efficient quantum circuit which computes $\QFT_{\cHp}$. First,
note that we are interested in a realization on a quantum computer
which works on qubits. This means that we have to embed the states and
transformations into a register of size $2^n$ for some positive
integer $n$. In
the following we will assume that $n$ is the smallest integer such
that $p < 2^n$ and we will identify the group elements $(x, y, z) \in
\cHp$ with a subset of the binary strings of length $3n$: in
each of the three components we choose the basis vectors $\ket{0},
\ldots, \ket{p-1}$ to represent the respective component of the
% changes MR, 03/10/05
element $(x,y,z)$. The following proposition shows that a $\QFT$ for
$\cHp$ can be implemented efficiently in terms of elementary quantum
gates.

\begin{proposition} 
\label{prop:qftcHp}
Let $p$ be prime, let $\cHp$ be the Heisenberg
group of order $p^3$ and let $\irr(\cHp) = \{ \chi_{a,b} : (a,b)
\in \F_p^2\} \cup \{ \rho_k : k = 1, \ldots, p-1 \}$ denote the
irreducible representations of $\cHp$. Then the $\QFT$ for $\cHp$ with
respect to $\irr(\cHp)$ can be computed using $O(\log^3 p)$
elementary quantum gates.
\end{proposition}
\begin{proof}
First we consider the normal subgroup $N_\infty \lhd \cHp$
and compute a Fourier transform for this abelian group. This group is
isomorphic to a direct product of two cyclic groups, i.\,e., $N_\infty
\cong \Z_p \times \Z_p$. The elements of $N_\infty$ are given by 
$N_\infty = \{ (0, y, z) : y, z \in \Z_p\}$, i.\,e., we can identify
the elements of $N_\infty$ with those binary strings of length $3n$
which have trivial support on the first $n$ positions. 
Note that the
irreducible representations of $N_\infty$ are given by $\psi_{a,b}$
for $a,b \in \Z_p$, where
\[
\psi_{a,b}(0,y,z) := \exp(2 \pi i/p (ay + bz)) = \omega_p^{ay+bz}. 
\]
Since $N_\infty$ is normal the group $\cHp$ operates on the
irreducible representations~\cite{CR:62}. 
We denote this action by ``$*$'', i.\,e., we have
a map 
$*: \cHp \times \irr(\cHp) \rightarrow \irr({\cal H}_p)$ 
which is explicitly given by 
$(x,y,z) * \psi_{a,b} = \psi_{a,b-ax}$. 
%
%Since the $\QFT$ for a direct product is given by the tensor product
%of the $\QFT$s of the components we have to apply the unitary
%transformation $\QFT_p \otimes \QFT_p$ in to compute
%$\QFT_{N_\infty}$. As an application of quantum phase 
%estimation it has
%been shown in \cite{Kitaev:97} that a $\QFT_{\Z_p}$ can be
%approximated using $O(\log^2 p)$ elementary gates. The circuit uses
%$O(\log p)$ ancilla qubits and reduces the computation of
%$\QFT_{\Z_p}$ to a computation of $\QFT_{\Z_{2^n}}$ for which
%efficient circuits are known \cite{Coppersmith:94}. 

Next, we choose as a transversal for $N_\infty \lhd \cHp$ the
ordered list $T=[(x, 0, 0) : x \in \Z_p]$. We have to be able to
efficiently implement the images of all irreducible
representations of $\cHp$
evaluated at the elements of $T$. This is required for the so-called
`twiddle factors' in the decomposition of $\QFT_{\cHp}$ along the
subgroup tower $\{1\} \lhd N_\infty \lhd \cHp$. 
Indeed, we construct a
$\QFT$ adapted to this subgroup tower,
see also \cite{PRB:99,MRR:2004}. We now use the
following formula for implementing a $\QFT_G$ which holds in the
situation where we have an abelian normal subgroup $N$ and an abelian
factor group $G/N$:
\[
\QFT_{\cHp} = 
\left(\onemat_{|G/N|} \otimes \QFT_N\right) \; 
\left(\bigoplus_{t \in T} \, \Phi(t) \right) \;
 \left(\QFT_{G/N} \otimes \onemat_{|N|}\right).
\]
Here $\Phi$ denotes an extension of the decomposition of the
regular representation of $N$ into irreducibles. Denoting this 
direct sum by $\Lambda$, i.\,e., 
$\Lambda := \bigoplus_{t \in T} \Phi(t)$, this means that we
have to implement the following transformation: 
\[
\Lambda : \ket{x} \ket{a} \ket{b} \mapsto 
\left\{
\begin{array}{lcl}
\ket{x} \ket{a} \ket{b-ax} & : & \mbox{if} \; a\not=0, \\
\omega_p^{xb} \ket{x} \ket{0} \ket{b} & : & \mbox{if} \; a=0.
\end{array} \right.
\]

It is straightforward to implement $\Lambda$ using classical efficient
circuits for modular addition and multiplication. Hence $\Lambda$ can
be implemented using $O(\log^3(p))$ quantum gates. Note that 
$\QFT_{N_\infty} = \QFT_{\Z_p} \otimes \QFT_{\Z_p}$ and 
$\QFT_{G/N_\infty} = \QFT_{\Z_p}$, both of which can be
either implemented approximately \cite{Kitaev:95} or exactly
\cite{MZ:2003} on a quantum computer using $O(\log^2 p)$ many
elementary quantum gates. Hence the claimed complexity for computing a
quantum Fourier transform for $\cHp$ follows. 
\end{proof}

%%%%%%%%%%%%%%%%%%%%%%%%%%%%%%%%%%%%%%%%%%%%%%%%%%%%%%%%%%%%
%
% Section: Random bases and Fourier sampling
%
%%%%%%%%%%%%%%%%%%%%%%%%%%%%%%%%%%%%%%%%%%%%%%%%%%%%%%%%%%%%

\section{Random bases and Fourier sampling}

\subsection{Nearly orthogonal vectors}
\label{subsec:nearlyortho}
In this subsection, we state some results about
sets of nearly orthogonal unit vectors in a Hilbert space.
We use $\norm{\cdot}$ to
denote the $\ell_2$-norm of vectors as
well as the $\ell_2$-induced operator norm of matrices. We
use $\totvar{v}$ to denote the $\ell_1$-norm of a vector $v$. We let
$\trnorm{M} = {\rm Tr} \sqrt{M^\dag M}$ denote the trace norm of 
a matrix $M$. For subspaces $V_1, V_2$ having trivial intersection,
their {\em overlap} is defined as 
$\ovlap{V_1}{V_2} = \max_{v_1, v_2} |\braket{v_1}{v_2}|$,
where $v_i$ range over unit vectors in $V_i$. Let $\Pi_i$ denote
the orthogonal projection operator onto $V_i$. It is easy to
see that $\ovlap{V_1}{V_2} = \norm{\Pi_1 \Pi_2}$.

\begin{proposition}
\label{prop:ovlap}
Let $V_1, V_2$ be subspaces of a Hilbert space having trivial
intersection. Let $\sigma_2$ denote the totally mixed state
in $V_2$. Let $V'_2$ denote the orthogonal complement of $V_1$
in $V_1 + V_2$ and $\sigma'_2$ denote the totally mixed
state in $V'_2$. Let $\delta = \ovlap{V_1}{V_2}$. Then,
\[
\trnorm{\sigma_2-\sigma'_2} \leq 2 \delta^{1/2} (1-\delta^2)^{-1/4}.
\]
\end{proposition}
\begin{proof}
Let $d = \dim V_2$ and $a_1, \ldots, a_d$ be an orthonormal
basis for $V_2$. Let $a'_1, \ldots, a'_d$ be the Gram-Schmidt
% changes MR, 03/10/05
orthonormalisation of $a_1, \ldots, a_d$ with respect to $V_1$. Hence,
$a'_1, \ldots, a'_d$ is an orthonormal basis for $V'_2$. We will
show that 
$\trnorm{\ketbra{a_i} - \ketbra{a'_i}} \leq 
2 \delta^{1/2} (1-\delta^2)^{-1/4}$ for all $1 \leq i \leq d$.
Since $\sigma_2 = \frac{1}{d} \sum_{i=1}^d \ketbra{a_i}$ and
$\sigma'_2 = \frac{1}{d} \sum_{i=1}^d \ketbra{a'_i}$, we will get
\[
\trnorm{\sigma_2-\sigma'_2} \leq
\frac{1}{d} \sum_{i=1}^d \trnorm{\ketbra{a_i} - \ketbra{a'_i}} \leq 
2 \delta^{1/2} (1-\delta^2)^{-1/4}.
\]

Fix some $i$, $1 \leq i \leq d$. Let $b_i + c_i$ denote the
orthogonal projection of $a_i$ onto the space spanned by
$V_1$ and $a_1, \ldots, a_{i-1}$, where $b_i \in V_1$ and
$c_i \in \linspan{a_1, \ldots, a_{i-1}} \subseteq V_2$. Then,
\begin{eqnarray*}
1 \geq \norm{b_i + c_i}^2 
& \geq & \norm{b_i}^2 + \norm{c_i}^2 - 2 |\braket{b_i}{c_i}| \\
& \geq & \norm{b_i}^2 + \norm{c_i}^2 - 
         2 \delta \norm{b_i} \norm{c_i} \\
&   =  & (1 - \delta^2) \norm{b_i}^2 + 
         (\delta \norm{b_i} -\norm{c_i})^2 \\
& \geq & (1 - \delta^2) \norm{b_i}^2, 
\end{eqnarray*}
i.\,e. $\norm{b_i} \leq \frac{1}{\sqrt{1 - \delta^2}}$. 
Now,
\[
\norm{b_i + c_i}^2 = \braket{a_i}{b_i + c_i} = 
\braket{a_i}{b_i} + \braket{a_i}{c_i} =
\braket{a_i}{b_i} \leq \frac{\delta}{\sqrt{1 - \delta^2}},
\]
i.\,e. $\norm{b_i + c_i} \leq \delta^{1/2} (1 - \delta^2)^{-1/4}$.
The third equality above follows from the fact that 
$a_1, \ldots, a_{i-1}, a_i$ are pairwise orthogonal. Now
$\braket{a_i}{a'_i} = \norm{a_i - b_i - c_i} = 
 \sqrt{1 - \norm{b_i + c_i}^2}$, and hence,
\[
\trnorm{\ketbra{a_i} - \ketbra{a'_i}} = 
2 \sqrt{1 - |\braket{a_i}{a'_i}|^2} =
2 \norm{b_i + c_i} \leq
2 \delta^{1/2} (1 - \delta^2)^{-1/4}.
\]

This completes the proof of the proposition.
\end{proof}
\begin{proposition}
\label{prop:gs}
Let $v'_1, \ldots, v'_n$ be unit vectors in a Hilbert space. Let
$0 \leq \delta < \frac{1}{2 n}$. Suppose
for all $i, j, i \neq j$, $|\braket{v_i}{v_j}| \leq \delta$.
Let $v_1, \ldots, v_n$ be unit vectors obtained by 
Gram-Schmidt orthonormalising $v'_1, \ldots, v'_n$. 
Then for any $i$, $1 \leq i \leq n$,
\[
\trnorm{\ketbra{v_i} - \ketbra{v'_i}} < 2 \sqrt{6} \cdot 
                                        \delta \sqrt{n}.
\]
\end{proposition}
\begin{proof}
Fix some $i$, $1 \leq i < n$. Let 
$a_{i+1} = \sum_{j=1}^i \alpha_j v'_j$ be the orthogonal projection
of $v'_{i+1}$ onto the subspace spanned by $v'_1, \ldots, v'_i$.
Then for all $k$, $1 \leq k \leq i$, 
$\braket{v'_k}{v'_{i+1} - a_{i+1}} = 0$ i\,.e.
$\braket{v'_k}{v'_{i+1}}=\sum_{j=1}^i \alpha_j \braket{v'_k}{v'_j}$.
Suppose $k$, $1 \leq k \leq i$ is such that 
$|\alpha_k| = \max_{j: 1 \leq j \leq i} |\alpha_j|$.
Then, 
\begin{eqnarray*}
\delta \geq |\braket{v'_k}{v'_{i+1}}| 
& \geq & |\alpha_k| |\braket{v'_k}{v'_k}| - 
         \sum_{\stackrel{j: 1 \leq j \leq i}{j \neq k}} 
         |\alpha_j| |\braket{v'_k}{v'_j}| \\
& \geq & |\alpha_k| - 
         \sum_{\stackrel{j: 1 \leq j \leq i}{j \neq k}} 
         |\alpha_k| |\braket{v'_k}{v'_j}| \\
& \geq & |\alpha_k| (1 - (i - 1) \delta)\\
&   >  & |\alpha_k| \cdot \frac{1}{2},
\end{eqnarray*}
i\,.e. $\max_{j: 1 \leq j \leq i} |\alpha_j| < 2 \delta$.
Now,
\begin{eqnarray*}
\norm{a_{i+1}}^2
& \leq & \sum_{j=1}^i |\alpha_j|^2 \norm{v_j}^2 +
         \sum_{\stackrel{j,j': 1 \leq j, j' \leq i}{j \neq j'}}
         |\alpha_j| |\alpha_{j'}| |\braket{v_j}{v_{j'}}| \\
&   <  & 4 \delta^2 n + 4 \delta^3 n^2 \\
&   <  & 4 \delta^2 n + 2 \delta^2 n \\
&   =  & 6 \delta^2 n.
\end{eqnarray*}
Reasoning as at the end of the proof of Proposition~\ref{prop:ovlap},
we get
\begin{eqnarray*}
\trnorm{\ketbra{v_{i+1}} - \ketbra{v'_{i+1}}} 
& = & 2 \sqrt{1 - |\braket{v_{i+1}}{v'_{i+1}}|^2} \\
& = & 2 \sqrt{1 - \norm{v'_{i+1} - a_{i+1}}^2} \\
& = & 2 \sqrt{1 - (1 - \norm{a_{i+1}}^2)} \\
& = & 2 \norm{a_{i+1}} \\
& < & 2 \sqrt{6} \cdot \delta \sqrt{n}.
\end{eqnarray*}
This completes the proof of the proposition.
\end{proof}

\subsection{Random orthonormal vectors}
\label{subsec:randorthonormal}
In this subsection, we state some facts about 
random orthonormal sets of vectors in $\C^d$.
One way of generating a random unit vector in $\C^d$ is
as follows: Consider $(y_1, \ldots, y_{2d}) \in \R^{2d}$, where
each $y_i$ is independently chosen according to the one dimensional
Gaussian distribution with mean $0$ and variance $1$ 
(i.\,e. $y_i$ is a
real valued random variable with probability density function 
$\frac{1}{\sqrt{2 \pi}} \exp(-y^2 / 2)$). Normalise to get the unit
vector $(x_1, \ldots, x_{2d})$, where 
$x_i = \frac{y_i}{\sqrt{y_1^2 + \cdots + y_{2d}^2}}$ (note that
any $y_i = 0$ with zero probability). We thus get a random
unit vector in $\R^{2d}$. Identifying a pair
of real numbers with a single complex number, we get a random unit
vector $(z_1, \ldots, z_d)$ in $\C^{d}$. To generate a random
orthonormal ordered set $\{v_1, \ldots, v_m\}$ of vectors in
$\C^d$, we can first sample $m$ unit vectors $\{v'_1, \ldots, v'_m\}$
in $\C^d$ and then do Gram-Schmidt orthonormalisation on them
to get $\{v_1, \ldots, v_m\}$ (note that with probability $1$,
$\{v'_1, \ldots, v'_m\}$ are linearly independent).

The following fact can be proved by combining Theorem~14.3.2 and 
Proposition~14.3.3 of \cite[Chapter 14]{matousek:dg} and using
the concavity of the square-root function.
\begin{fact}
\label{fact:1dimproj}
Let $t > 0$, and
$\ket{v}$, $\ket{w}$ independent random unit vectors in
$\C^d$. Then,
\[
\Pr\left[|\braket{v}{w}| > t + \frac{10}{\sqrt{d}}\right] 
 \leq 2 \exp(-t^2 d).
\]
\end{fact}

We will require the following upper and lower bounds on the
tails of the chi-square distribution (the chi-square distribution
with $d$ degrees of freedom is the sum of squares of $d$ independent
Gaussians with mean $0$ and variance $1$). The
upper bound can be proved via Chernoff-style arguments on the
moment generating function of the chi-square distribution. 
The lower bound follows, for example, from the central limit 
theorem in
probability theory. One can also give a direct proof of the lower 
bound using the
probability density function of the chi-square distribution 
and estimating
it via Stirling's approximation of the gamma function.
\begin{fact}
\label{fact:chisq}
Let $(X_1, \ldots, X_d)$ be independent random 
variables such that $X_i$ is one-dimensional
Gaussian with mean $0$ and variance $1$.
Let $X^2 = X_1^2 + \cdots + X_d^2$. Let $0 \leq \epsilon < 1/2$.
There exists a universal constant $\gamma > 0$ such that
\begin{enumerate} 
\item $\Pr[|X^2 - d| > d \epsilon] < 2 \exp(-d \epsilon^2 / 6)$, 
\item $\Pr[X^2 > d + \sqrt{d}] > \gamma$, 
      $\Pr[X^2 < d - \sqrt{d}] > \gamma$. 
\end{enumerate}
\end{fact}
The following result follows easily from Fact~\ref{fact:chisq}.
% changes MR, 03/10/05
% changes PS, 03/10/05
A similar result appears as Lemma~2 in \cite{MRRS:2004}.
\begin{fact}
\label{fact:randspace}
Let $V = \{a^1, \ldots, a^p\}$ be a random orthonormal set of
$p$ vectors in $\C^d$. Let $a^i_j$ denote the $j$th
coordinate of vector $a^i$.
Define the $d$-dimensional probability
vector $S$ as follows: $S_j = \frac{1}{p} \sum_{i=1}^p |a^i_j|^2$.
Let $0 \leq \epsilon < 1/2$. 
Suppose $p = \Omega(\epsilon^{-2} \log d)$.
Let $U$ denote the uniform probability distribution on 
$\{1, \ldots, d\}$.
Then, with probability at least $1 - \exp(-\Omega(\epsilon^2 p))$
over the choice of $V$, $\totvar{S - U} \leq \epsilon$.
\end{fact}

We will also need the following Chernoff upper bounds on the tail 
of the sum of $d$ independent identically distributed binary 
random variables.
\begin{fact}[\mbox{\cite[Cor. A.7, 
                         Theorem A.13]{AlSp:probmethod}}]
\label{fact:chernoff}
Let $(X_1, \ldots, X_d)$ be independent binary random 
variables such that $\Pr[X_i = 1] = p$. Let $X = X_1 + \cdots + X_d$.
Let $0 \leq \epsilon < 1/2$.
Then, 
\begin{enumerate}
\item $\Pr\left[\left|\frac{X}{d} - p\right| > \epsilon\right] < 
       2 \exp(-2 \epsilon^2 d)$,
\item $\Pr[X < \frac{d p}{2}] < \exp(-d p / 8)$. 
\end{enumerate}
\end{fact}

\subsection{Hidden subgroup problem and Fourier sampling}
\label{subsec:hspintro}
In this subsection, we recall the standard approach to solving the
hidden subgroup problem based on Fourier sampling.
A $d$-dimensional unitary representation of $G$ is a 
group homomorphism 
$\rho: G \rightarrow \U(d)$, where $\U(d)$ is the group of 
$d \times d$
complex unitary matrices under multiplication. 
Let $\C[G]$ denote the group algebra;
it is a $|G|$-dimensional Hilbert space over $\C$
with group elements $\ket{g}$, $g \in G$ as an orthonormal basis.
Let ${\cal R}[G]$ denote the $|G|$-dimensional Hilbert space
over $\C$ spanned by the orthonormal basis vectors 
$\ket{\rho, i, j}$, where $\rho$ runs over inequivalent
irreducible unitary representations of $G$
and $i, j$ run over the row and column indices of $\rho$.
The quantum Fourier transform over $G$, $\QFT_G$, is the 
following $\C$-linear map from $\C[G]$ to ${\cal R}[G]$ defined 
as follows: 
\[
\ket{g} \mapsto \sum_{\rho} \sqrt{\frac{d_\rho}{|G|}} 
\sum_{i, j = 1}^{d_\rho} \rho_{i j}(g) \ket{\rho, i, j},
\]
where $d_\rho$ denotes the dimension of $\rho$. $\QFT_G$ is an
inner product preserving map from $\C[G]$ to ${\cal R}[G]$.

For a subset $T \subseteq G$,
define $\ket{T} = \frac{1}{\sqrt{|T|}} \sum_{t \in T} \ket{t}$ to
be the uniform superposition over elements of $T$. For
a representation $\rho$, define the matrix 
$\rho(T) = \frac{1}{\sqrt{|T|}} \sum_{t \in T} \rho(t)$.
If $H \leq G$, it can be shown (see e.g.~\cite{HRT:2003})
that $\frac{1}{\sqrt{|H|}} \rho(H)$
is an orthogonal projection onto the subspace $V_H^\rho$ of the
representation space of $\rho$ consisting of all vectors
$\ket{v}$ such that $\rho(h) \ket{v} = \ket{v}$ for all $h \in H$.
Thus, $\rank{\rho(H)} = \dim V_H^\rho$.

In the strong Fourier sampling method for the hidden subgroup
problem, we begin by forming the uniform
superposition $\frac{1}{\sqrt{|G|}} \sum_{g \in G} \ket{g}\ket{0}$ 
and then query $f$ to get the superposition 
$\frac{1}{\sqrt{|G|}} \sum_{g \in G} \ket{g} \ket{f(g)}$. We then
measure the second register to get a
uniform mixture over vectors $\ket{g H}$ in the first register.
Assuming the first
register is in state $\ket{g H}$ for some particular $g \in G$,
its state after the application of $\QFT_G$ becomes
\[
\frac{1}{\sqrt{|G| |H|}} \sum_{\rho, i, j} \sqrt{d_\rho}
 \sum_{h \in H} \rho_{i j}(g h) \ket{\rho, i, j}.
\]
If we now measure the representation name and column index, we 
sample $(\rho, j)$ with probability
\[
P^G_{H}(\rho, j) = \frac{d_\rho}{|G|} \sum_i |\rho_{ij}(g H)|^2 =
\frac{d_\rho}{|G|} \norm{\rho(g H) \ket{j}}^2 =
\frac{d_\rho}{|G|} \norm{\rho(H) \ket{j}}^2.
\]
The third equality above follows from the fact that
$\norm{\rho(g H) \ket{j}} = \norm{\rho(g) \rho(H) \ket{j}} =
 \norm{\rho(H) \ket{j}}$, since $\rho(g)$ is unitary. Thus, 
as long as we measure just the representation name and column
index $(\rho, j)$, the probabilities are independent of the actual
coset $g H$ that we find ourselves in. This fact can be viewed
as the non-abelian generalisation of the fact that in abelian
Fourier sampling the probability distribution on the characters
is independent of the actual coset that we land up in.
Also, it can be shown that (see \cite{GSVV:2004})
\[P^G_{H}(\rho) = 
 \sum_{j = 1}^{d_\rho} \frac{d_\rho}{|G|} \norm{\rho(H) \ket{j}}^2 =
 \frac{d_\rho |H|}{|G|} \rank{\rho(H)} =
 \frac{d_\rho |H|}{|G|} \dim V^\rho_H.
\]
In {\em weak Fourier sampling}, we only measure the names 
$\rho$ of the representations and ignore the column indices $j$.
It can be shown (see e.g.~\cite{HRT:2003}) that for normal
hidden subgroups $H$, no more information about $H$ is contained 
in the
column space of the resulting state after the measurement of
$\rho$. Thus, weak Fourier sampling is the optimal measurement
to recover a normal hidden subgroup starting from the uniform mixture
of coset states.

Define a distance measure
$w(G; H_1, H_2) = \sum_\rho |P^G_{H_1}(\rho) - P^G_{H_2}(\rho)|$
between subgroups $H_1, H_2 \leq G$.
$w(G; H_1, H_2)$ is the total variation distance between the
probability distributions, when the hidden subgroup is $H_1$ or
$H_2$, on the names of the representations
obtained via weak Fourier sampling. 
\cite{HRT:2003, GSVV:2004}
show that $O(\log |G|)$ weak Fourier samplings suffice to 
reconstruct the {\em normal core}
$\core{H}$ of the hidden subgroup $H$, 
where $\core{H}$ is the largest normal subgroup of 
$G$ contained in $H$. 
%\cite{KS:2004} gives a lower bound
%on $w(G; H, \{1\})$ ($\{1\}$ denotes the identity subgroup)
%in terms of the intersections sizes of the conjugacy classes of
%$G$ with $H$. 
Adapting their arguments, we prove the following result.
\begin{proposition}
\label{prop:normcore}
Let $H_1, H_2 \leq G$. Suppose $\core{H_1} \neq \core{H_2}$. 
Then, $w(G; H_1, H_2) \geq 1/2$.
\end{proposition}
\begin{proof}
Let $N_1 = \core{H_1}$ and $N_2 = \core{H_2}$. Without loss of
generality, $N_1 \not \leq N_2$. Define the kernel of a 
representation 
$\ker{\rho} = \{g \in G: \rho(g) = \onemat_{d_\rho}\}$;
$\ker(\rho) \lhd G$.
It can be shown 
(see e.g.~\cite{HRT:2003}) for an irreducible representation $\rho$
and a subgroup $H \leq G$,
that if $\rank{\rho(H)} > 0$, $\core{H} \lhd \, \ker{\rho}$. 
Hence,
\[
1 = \sum_\rho P^G_{H_2}(\rho) =
    \sum_\rho \frac{d_\rho |H_2|}{|G|} \cdot \rank{\rho{H_2}} =
    \sum_{\rho: N_2 \lhd \, \ker{\rho}} \frac{d_\rho |H_2|}{|G|}
    \cdot \rank{\rho(H_2)}.
\]
Since $N_1 \lhd G$, $N_1 H_2$ is a subgroup of $G$. Hence,
$N_1 \lhd \core{N_1 H_2}$ and $N_2 \lhd \core{N_1 H_2}$. 
Since $N_1 \not \leq H_2$, $|N_1 H_2| \geq 2 \cdot |H_2|$.
For an
irreducible representation $\rho$ such that $N_1 \lhd \, \ker{\rho}$,
\[
\rank{\rho(N_1 H_2)} = \rank{\rho(N_1) \rho(H_2)} = 
\rank{\rho(H_2)}.
\] 
Also,
\[
1 = \sum_{\rho: N_1, N_2 \lhd \, \ker{\rho}} 
    \frac{d_\rho |N_1 H_2|}{|G|} \cdot \rank{\rho(N_1 H_2)} \geq 
    2 \cdot \sum_{\rho: N_1, N_2 \lhd \, \ker{\rho}} 
    \frac{d_\rho |H_2|}{|G|} \cdot \rank{\rho(H_2)},
\]
i\,.e. 
\[
\sum_{\rho: N_1, N_2 \lhd \, \ker{\rho}} \frac{d_\rho |H_2|}{|G|}
\cdot \rank{\rho(H_2)} \leq \frac{1}{2}.
\]
Finally,
\begin{eqnarray*}
w(G; H_1, H_2)
&   =  & \sum_\rho \frac{d_\rho}{|G|} \cdot
        ||H_1| \, \rank{\rho(H_1)} - |H_2| \, \rank{\rho(H_2)}| \\
& \geq & \sum_{\rho: N_2 \lhd \, \ker{\rho},
                     N_1 \not \lhd \, \ker{\rho}}
         \frac{d_\rho}{|G|} \cdot
         |H_2| \, \rank{\rho(H_2)} \\
& \geq & 1 - \sum_{\rho: N_1, N_2 \lhd \, \ker{\rho}} 
             \frac{d_\rho}{|G|} \cdot |H_2| \, \rank{\rho(H_2)} \\
& \geq & \frac{1}{2}.
\end{eqnarray*}
This completes the proof.
\end{proof}

For a normal subgroup $N \lhd G$, define the {\em normal core family}
of $N$, $\corefamily{N} = \{H: H \leq G, \core{H} = N\}$. In view
of Proposition~\ref{prop:normcore}, the remaining 
challenge is to distinguish between subgroups $H_1, H_2$ from the
same normal core family.
%\begin{fact}
%Let $C_1, \ldots, C_k$ denote the conjugacy classes of $G$. 
%Let $H_1 \,\triangle\, H_2$ denote the symmetric difference, as sets,
%of $H_1$ and $H_2$.
%Then, 
%\[
%|H_1 \,\triangle\, H_2|^{-1} \sum_{i = 1}^k 
%(|C_i \cap H_1| - |C_i \cap H_2|)^2 |C_i|^{-1} \leq 
%w(G; H_1, H_2) \leq
%\sum_{i = 1}^k |C_i \cap (H_1 \,\triangle\, H_2)| |C_i|^{-1/2}.
%\]
%\end{fact}

The success of strong Fourier sampling depends on how much statistical
information about $H$ is present in the probability distribution
$P^G_H(\rho, j)$. The amount of information, in general, depends on
the choice of basis for each representation $\rho$, i.\,e., on the
choice of basis for $j$; see \cite{MRRS:2004} for more details.
Grigni et al.~\cite{GSVV:2004} show that under certain conditions on
$G$ and $H$, the {\em random strong} Fourier sampling method, where a
random choice of basis is made for each representation, gives
exponentially small information about distinguishing $H$ from the
identity subgroup.  In the next section, we prove a complementary
result viz. under different conditions on $G$, $(\log |G|)^{O(1)}$
random strong Fourier samplings do give enough information to
reconstruct the hidden subgroup $H$ with high probability.

%%%%%%%%%%%%%%%%%%%%%%%%%%%%%%%%%%%%%%%%%%%%%%%%%%%%%%%%%%%%
%
% Section: Power of the random strong method
%
%%%%%%%%%%%%%%%%%%%%%%%%%%%%%%%%%%%%%%%%%%%%%%%%%%%%%%%%%%%%

\section{Power of the random strong method}
\label{sec:power}
In this section, we define a parameter $r(G)$ on a
group $G$ which, if at least  $(\log |G|)^{-O(1)}$, suffices
for the random strong method to identify the hidden subgroup
with $(\log |G|)^{O(1)}$ Fourier samplings. Let $H_1, H_2 \leq G$.
We first define a distance measure $r(G; H_1, H_2)$ 
between $H_1, H_2$. In what follows, we use the notation of
Section~\ref{subsec:hspintro}.
\begin{definition}[$r(G; H_1, H_2; \rho)$]
\label{def:rh1h2}
Suppose $\rho$
is an irreducible $d_\rho$-dimensional unitary representation of $G$.
Let $\Pi_i$ denote the orthogonal projection onto $V^\rho_{H_i}$ i.\,e.
$\Pi_i = \frac{1}{|H_i|} \sum_{h \in H_i} \rho(h)$.
Let $\Pi_{1,2}$ denote the orthogonal projection onto
$V^\rho_{H_1} \cap V^\rho_{H_2}$. It is easy to check that
$V^\rho_{H_1} \cap V^\rho_{H_2} = V^\rho_{\left<H_1, H_2\right>}$,
where $\left<H_1, H_2\right>$ denotes
the subgroup of $G$ generated by $H_1$ and $H_2$. Thus,
$\Pi_{1,2} = \frac{1}{|\left<H_1, H_2\right>|} 
             \sum_{h \in \left<H_1, H_2\right>} \rho(h)$.
Define $\Pi'_i = \Pi_i - \Pi_{1,2}$. 
$\Pi'_i$ is the orthogonal projection
onto the subspace $V'_i$ defined as the orthogonal complement of
$V^\rho_{H_1} \cap V^\rho_{H_2}$ in $V^\rho_{H_i}$. $V'_1$ and
$V'_2$ have trivial intersection.
Define $r_i = \rank{\Pi_i}$ and $r'_i = \rank{\Pi'_i}$. 
Define $\hat{h} = \max\{|H_1| r_1, |H_2| r_2\}$, 
$\tilde{h} = |(|H_1| r_1 - |H_2| r_2)|$ and 
$\delta = \norm{\Pi'_1 \Pi'_2}$. Recall that 
$\delta = \ovlap{V'_1}{V'_2}$.
Consider the following three cases:
\begin{enumerate}
\item When $\frac{\sqrt{d_\rho}}{\log |G|} = 
      \Omega((r_1 + r_2)^{3/2})$. Loosely speaking,
      $r_1, r_2$ are both small. In this case, define
      \[
      r(G; H_1, H_2; \rho) = 
      \max\left\{
      \frac{\hat{h}}{2} \left(
      \Omega\left(
      \frac{\sqrt{r'_1}}{r_1} + \frac{\sqrt{r'_2}}{r_2}
      \right) - 
      2 \delta^{1/2} (1 - \delta^2)^{-1/4}
      \right), \tilde{h}
      \right\}.
      \]

\item When $\frac{\sqrt{d_\rho}}{\log |G|} = \Omega(r_1)$ and
      $\frac{r_2}{r_1} = \Omega(\log^2 |G|)$. Loosely speaking,
      $r_1$ is small
      and $r_2$ is relatively large with respect to $r_1$.
      In this case, define
      \[
      r(G; H_1, H_2; \rho) = 
      \max\left\{
      \frac{\hat{h}}{2} \cdot
      \Omega\left(\frac{1}{\sqrt{r_1}}\right), \tilde{h}
      \right\}.
      \]

\item Otherwise, define 
      $r(G; H_1, H_2; \rho) = \tilde{h}$.
\end{enumerate}
\end{definition}
\begin{definition}[$r(G; H_1, H_2)$, $r(G)$]
Let $H_1, H_2 \leq G$. Define
$r(G; H_1, H_2) = \sum_\rho \frac{d_\rho}{|G|} \cdot
                            r(G; H_1, H_2; \rho)$ 
and $r(G) = \min_{H_1, H_2} r(G; H_1, H_2)$. 
\end{definition}
From the above definition, it is easy to see that
$r(G; H_1, H_2) \geq w(G; H_1, H_2)$.
\begin{definition}[$P^G_{H, \cB}$]
Let $\cB$ be a set of orthonormal bases for the irreducible
unitary representations of $G$. Suppose $H \leq G$.
$P^{G, \cB}_H$ denotes the probability distribution on the
representation names and column indices $(\rho, j)$ got by
strong Fourier sampling the state $\ket{H}$ according to $\cB$.
\end{definition}

The significance of $r(G; H_1, H_2)$ arises from the following
theorem.
\begin{theorem}
\label{thm:randbasis}
With probability at least $1 - \exp(-\Omega(\log^2 |G|))$ 
over the choice of random representation bases $\cB$
for Fourier sampling, 
\[
\trnorm{P^{G, \cB}_{H_1} - P^{G, \cB}_{H_2}} \geq 
r(G; H_1, H_2).
\]
\end{theorem}
Using this theorem, we can
apply a `minimum-finding-like' algorithm to identify the hidden
subgroup.
\begin{corollary}
\label{cor:randbasis}
Let $s(G)$ denote the number of distinct subgroups of $G$.
With probability at least $2/3$
% changes MR, 03/10/05
% changes PS, 03/10/05
over the choice of random bases for representations of $G$,
Fourier sampling 
$O\left(\frac{\log s(G)}{r^2(G)}\right)$ times in a random basis
gives enough classical information
to identify a hidden subgroup in $G$. In particular,
$O\left(\left(\frac{\log |G|}{r(G)}\right)^2\right)$ random
strong Fourier samplings suffice.
\end{corollary}
\begin{proof}
From Theorem~\ref{thm:randbasis}, we get that
for all pairs of subgroups $H_1, H_2 \leq G$,
with probability at least 
$1 - \exp(-\Omega(\log^2 |G|))$
over the choice of random bases $\cB$ for representations of $G$, 
$\trnorm{P^{G, \cB}_{H_1} - P^{G, \cB}_{H_2}} \geq r(G)$.
Call a set of representation bases $\cB$ {\em good} if
$\trnorm{P^{G, \cB}_{H_1} - P^{G, \cB}_{H_2}} \geq r(G)$
for all pairs of subgroups $H_1, H_2 \leq G$.
By the union bound on probabilities, 
a random choice of representation bases gives a good $\cB$
with probability at least 
$1 - s(G) \exp(-\Omega(\log^2 |G|)) = 1 - \exp(-\Omega(\log^2 |G|))$.
Suppose we have such a good $\cB$. 
Under the promise that the hidden subgroup is
either $H_1$ or $H_2$, $\cB$ recognises which one it is
with confidence at least $1/2 + r(G)/4$ using Bayes's 
rule. Using Fact~\ref{fact:chernoff}, the confidence
can be boosted to at least $1 - \frac{1}{4 s(G)}$ by
% changes MR, 03/10/05
% changes PS, 03/10/05
Fourier sampling 
$O\left(\frac{\log s(G)}{r^2(G)}\right)$ times with $\cB$.
We can now run a classical `minimum-finding-like' algorithm on 
the measured samples, comparing two subgroups 
$H_1, H_2 \leq G$ at a time, to discover the actual hidden subgroup
$H$ in $G$ with confidence at least 
$1 - \frac{s(G)}{4 s(G)} = 3/4$. The overall confidence bound
becomes 
$(1 - \exp(-\Omega(\log^2 |G|))) \cdot \frac{3}{4} \geq \frac{2}{3}$.
The second bound follows from the fact that
$s(G) \leq 2^{\log^2 |G|}$,
since any group of size $a$ has at most $\log a$ generators.
\end{proof}

The rest of the section is devoted to proving 
Theorem~\ref{thm:randbasis}. We first prove some necessary
technical lemmas.
\begin{lemma}
\label{lem:randspace1}
Let $W = \{a^1, \ldots, a^p\} \cup \{b^1, \ldots, b^q\} \cup 
         \{c^1, \ldots, c^r\}$ be a random orthonormal set of
$p + q + r$ vectors in $\C^d$. Let $a^i_j$ denote the $j$th
coordinate of vector $a^i$; similar notations will be used for
the vectors $b^i$, $c^i$ too. Define two $d$-dimensional probability
vectors $S, T$ as follows: 
\[
S_j = \frac{1}{p + r} 
      \left(\sum_{i=1}^p |a^i_j|^2 + \sum_{i=1}^r |c^i_j|^2\right),
T_j = \frac{1}{q + r} 
      \left(\sum_{i=1}^q |b^i_j|^2 + \sum_{i=1}^r |c^i_j|^2\right).
\]
Then there exists $\delta = \theta((p + q + r)^{-3/2})$ such that
the following holds: Define
$\alpha = d \delta^2 - 2 \log (p + q + r)$. Suppose 
$\alpha = \Omega(1)$.
Then, with probability at least $1 - \exp(-\Omega(\alpha))$
over the choice of $W$, 
\[
\totvar{S - T} = \Omega\left(
 \frac{\sqrt{p}}{p + r} + \frac{\sqrt{q}}{q + r}
 \right).
\]
\end{lemma}
\begin{proof}
Generate a set 
$W' = \{a'^1, \ldots, a'^p\} \cup \{b'^1, \ldots, b'^q\} \cup 
      \{c'^1, \ldots, c'^r\}$ 
of $p + q + r$ random independent unit vectors in $\C^d$ 
as described in Section~\ref{subsec:randorthonormal}. 
Let $\alpha'^i_j$, $\beta'^i_j$, $\gamma'^i_j$, 
$j = \{1, \ldots, d\}$
denote the Gaussians used to generate the random unit vectors
$a'^i$, $b'^i$, $c'^i$ respectively. Then,
\[
a'^i_j = \frac{\alpha'^i_j}{\sum_{l = 1}^d |\alpha'^i_l|^2}, ~~~
b'^i_j = \frac{\beta'^i_j}{\sum_{l = 1}^d |\beta'^i_l|^2}, ~~~
c'^i_j = \frac{\gamma'^i_j}{\sum_{l = 1}^d |\gamma'^i_l|^2}.
\]

By Fact~\ref{fact:chisq}, with probability at least
$1 - 2 \exp(-d \delta^2)$ over the choice of the Gaussians, 
the normalisation factor in the denominator
of a given vector in $W'$ is $\sqrt{(1 \pm \epsilon) d}$ where
$\epsilon = O(\delta)$. 
Let $E_0$ be the event that
the normalisation factors in the denominators of all
vectors in $W'$ are $\sqrt{(1 \pm \epsilon) d}$.
By the union bound on probabilities, $E_0$ occurs 
with probability at least $1 - 2 \exp(-\alpha)$ over the choice of
the Gaussians.

Since $\alpha = \Omega(1)$, $\delta > \frac{10}{\sqrt{d}}$. 
By Fact~\ref{fact:1dimproj}, for any $w'^i, w'^j \in W'$, 
$i \neq j$, 
$
|\braket{w'^i}{w'^j}| \leq 
\delta + \frac{10}{\sqrt{d}} <
2 \delta
$
with probability at least $1 - 2 \exp(-d \delta^2)$ over the 
choice of the Gaussians.
Let $E_1$ denote the event that
$|\braket{w'^i}{w'^j}| < 2 \delta$ for all $w'^i, w'^j \in W'$,
$i \neq j$. 
By the union bound on probabilities, $E_1$ occurs
with probability at least $1 - 2 \exp(-\alpha)$ over the 
choice of the Gaussians.

Using Fact~\ref{fact:chisq} we see that for any fixed
coordinate $j$, with constant probability at least $\theta$ over the
choice of the Gaussians, each
of the following three events occurs\,:
\[
\sum_{i=1}^p |\alpha'^i_j|^2 > p + \sqrt{p}, ~~~
\sum_{i=1}^q |\beta'^i_j|^2 < q - \sqrt{q}, ~~~
\sum_{i=1}^r |\gamma'^i_j|^2 > r + 1. 
\]
Since these are independent events,
all three of them hold at coordinate $j$
simultaneously with constant probability at least
$\theta^3$. Call such a coordinate $j$ {\em good}.
Let $E_2$ denote the event that
more than $\frac{d \theta^3}{2}$ coordinates $j$ are good.
By Fact~\ref{fact:chernoff}, $E_2$ occurs with probability at least
$1 - \exp(d \theta^3 / 8)$ over the choice of the Gaussians. 

Now suppose that all three events $E_0, E_1, E_2$ occur.
We Gram-Schmidt orthonormalise $W'$ to get the random orthonormal
set 
$W = \{a^1, \ldots, a^p\} \cup \{b^1, \ldots, b^q\} \cup 
     \{c^1, \ldots, c^r\}$.
Let $S', T'$ be the analogous probability
vectors defined with respect
to $W'$ instead of $W$. 
From Proposition~\ref{prop:gs}, we see
that 
\[
\trnorm{\ketbra{w} - \ketbra{w'}} <
20 \cdot \delta \cdot (p + q + r) =
O((p + q + r)^{-1})
\]
for corresponding vectors $w \in W$, $w' \in W'$. Define
density matrices 
\[
\sigma  = \frac{1}{p + r} \left(
          \sum_{i=1}^p \ketbra{a^i} + \sum_{i=1}^r \ketbra{c^i}
          \right), ~~~
\sigma' = \frac{1}{p + r} \left(
          \sum_{i=1}^p \ketbra{a'^i} + \sum_{i=1}^r \ketbra{c'^i}
          \right),
\]
\[
\tau    = \frac{1}{q + r} \left(
          \sum_{i=1}^q \ketbra{b^i} + \sum_{i=1}^r \ketbra{c^i}
          \right), ~~~
\tau'   = \frac{1}{q + r} \left(
          \sum_{i=1}^q \ketbra{b'^i} + \sum_{i=1}^r \ketbra{c'^i}
          \right).
\]
Then, $S, S', T, T'$ are the probability distributions got by
measuring the states $\sigma, \sigma', \tau, \tau'$ in the
standard basis of $\C^d$. By triangle inequality,
$\trnorm{\sigma - \sigma'} = O((p + q + r)^{-1})$ and
$\trnorm{\tau - \tau'} = O((p + q + r)^{-1})$. Hence,
$\totvar{S - S'} = O((p + q + r)^{-1})$ and
$\totvar{T - T'} = O((p + q + r)^{-1})$. 

For a {\em good} coordinate $j$
\begin{eqnarray*}
|S'_j - T'_j| 
& > & \frac{p}{(p + r) (1 + \epsilon) d} -
      \frac{q}{(q + r) (1 - \epsilon) d} +
      \frac{r + 1}{(1 + \epsilon) d}
      \left(\frac{1}{p + r} - \frac{1}{q + r}\right) \\
&   & +\, \frac{\sqrt{p}}{(p + r) (1 + \epsilon) d} 
      +   \frac{\sqrt{q}}{(q + r) (1 - \epsilon) d} \\
& = & \frac{-2 q \epsilon}{(q + r) (1 - \epsilon^2) d} +
      \frac{q - p}{(p + r) (q + r) (1 + \epsilon) d} +
      \frac{\sqrt{p}}{(p + r) (1 + \epsilon) d} +
      \frac{\sqrt{q}}{(q + r) (1 - \epsilon) d} \\
& > & -O\left(\frac{1}{(p + q + r)^{1/2} (q + r) 
                       (1 - O((p + q + r)^{-3})) d}
        \right) + 
       \Omega\left(\frac{1}{d}
       \left(\frac{\sqrt{p}}{p + r}+\frac{\sqrt{q}}{q + r}\right)
       \right) \\
& = & \Omega\left(\frac{1}{d}
      \left(\frac{\sqrt{p}}{p + r} + \frac{\sqrt{q}}{q + r}\right)
      \right).
\end{eqnarray*}
The first, third and fourth steps above follow from the fact that 
$\epsilon = O(\delta) = O((p + q + r)^{-3/2})$ and $p \leq q$ without
loss of generality.

Now,
\[
\totvar{S' - T'} \geq
\sum_{j: j \mbox{ good}} |S'_j - T'_j| >
\frac{d \theta^3}{2} \cdot 
\Omega\left(\frac{1}{d}
\left(\frac{\sqrt{p}}{p + r} + \frac{\sqrt{q}}{q + r}\right)
\right) =
\Omega\left(
\frac{\sqrt{p}}{p + r} + \frac{\sqrt{q}}{q + r}
\right).
\]
Finally,
\begin{eqnarray*}
\totvar{S - T} 
& \geq & \totvar{S' - T'} - \totvar{S - S'} - \totvar{T - T'} \\
&   =  & \Omega\left(\frac{\sqrt{p}}{p + r} + \frac{\sqrt{q}}{q + r}
         \right) - 2 \cdot O\left(\frac{1}{p + q + r}\right) \\
&   =  & \Omega\left(\frac{\sqrt{p}}{p + r} + \frac{\sqrt{q}}{q + r}
         \right).
\end{eqnarray*}
The confidence bound is 
\[
\Pr[E_0 \wedge E_1 \wedge E_2] >
1 - 4 \exp(-\alpha) - \exp(d \theta^3 / 8) = 
1 - \exp(-\Omega(\alpha)),
\]
since $\delta = O(1)$.
This completes the proof of the lemma. 
\end{proof}

We can prove the following lemma in a similar fashion
as Lemma~\ref{lem:randspace1}.
\begin{lemma}
\label{lem:randspace2}
Let $W = \{a^1, \ldots, a^p\}$ be a random orthonormal set of
$p$ vectors in $\C^d$. Let $a^i_j$ denote the $j$th
coordinate of vector $a^i$.
Define the $d$-dimensional probability
vector $S$ as follows: $S_j = \frac{1}{p} \sum_{i=1}^p |a^i_j|^2$.
Then there exists $\delta = \theta(p^{-1})$ such that
the following holds: Define
$\alpha = d \delta^2 - 2 \log p$. Suppose 
$\alpha = \Omega(1)$.
Let $U$ denote the uniform probability distribution on
$\{1, \ldots, d\}$.
Then, with probability at least $1 - \exp(-\Omega(\alpha))$
over the choice of $V$, $\totvar{S - U} = \Omega(p^{-1/2})$.
\end{lemma}
\begin{proof} {\bf (Sketch)}
Generate a set $W' = \{a'^1, \ldots, a'^p\}$
of $p$ random independent unit vectors in $\C^d$ 
as described in Section~\ref{subsec:randorthonormal}. 
Let $\alpha'^i_j$, $j = \{1, \ldots, d\}$
denote the Gaussians used to generate the random unit vectors
$a'^i$. Then,
$a'^i_j = \frac{\alpha'^i_j}{\sum_{l = 1}^d |\alpha'^i_l|^2}$. 
We Gram-Schmidt orthonormalise $W'$ to get the random orthonormal
set $W = \{a^1, \ldots, a^p\}$. 
Let $E_0$ be the event that
the normalisation factors in the denominators of all
vectors in $W'$ are $\sqrt{(1 \pm \epsilon) d}$, where
$\epsilon = O(\delta)$.
$E_0$ occurs 
with probability at least $1 - 2 \exp(-\alpha))$ over the choice of
the Gaussians.
Let $E_1$ denote the event that
$|\braket{w'^i}{w'^j}| < 2 \delta$ for all $w'^i, w'^j \in W'$,
$i \neq j$. 
$E_1$ occurs
with probability at least $1 - 2 \exp(-\alpha)$ over the 
choice of the Gaussians.
Call a coordinate $j$ {\em good} if
$\sum_{i=1}^p |\alpha'^i_j|^2 > p + \sqrt{p}$.
Let $E_2$ denote the event that
more than $\frac{d \theta}{2}$ coordinates $j$ are good.
$E_2$ occurs with probability at least
$1 - \exp(d \theta / 8)$ over the choice of the Gaussians. 

Now suppose that all three events $E_0, E_1, E_2$ occur.
Let $S'$ be the analogous probability
vector defined with respect
to $W'$ instead of $W$. 
Then, $\totvar{S - S'} = O(p^{-1/2})$.
For a {\em good} coordinate $j$
\begin{eqnarray*}
\left|S'_j - \frac{1}{d}\right| 
& = & \frac{1}{(1 + \epsilon) d} -
      \frac{1}{d} + \frac{1}{\sqrt{p} (1 + \epsilon) d} \\
& = & \frac{-\epsilon}{(1 + \epsilon) d} +
      \frac{1}{\sqrt{p} (1 + \epsilon) d} \\
& = & -O\left(\frac{-1}{d p}\right) +
      \Omega\left(\frac{1}{d \sqrt{p}}\right) \\
& = & \Omega\left(\frac{1}{d \sqrt{p}}\right).
\end{eqnarray*}
The third step above follows from the fact that 
$\epsilon = O(\delta) = O(p^{-1})$.
Hence,
\[
\totvar{S' - U} \geq
\sum_{j: j \mbox{ good}} |S'_j - T'_j| >
\frac{d \theta}{2} \cdot 
\Omega\left(\frac{1}{d \sqrt{p}}\right) =
\Omega\left(\frac{1}{\sqrt{p}}\right).
\]
Finally,
\[
\totvar{S - U} 
 \geq \totvar{S' - U} - \totvar{S - S'} 
   =  \Omega(p^{-1/2}) - O(p^{-1}) 
   =  \Omega(p^{-1/2}).
\]
The confidence bound is 
\[
\Pr[E_0 \wedge E_1 \wedge E_2] >
1 - 4 \exp(-\alpha) - \exp(d \theta / 8) = 
1 - \exp(-\Omega(\alpha)),
\]
since $\delta = O(1)$.
This completes the proof of the lemma. 
\end{proof}

We are now in a position to finally prove Theorem~\ref{thm:randbasis}.

\begin{proof} {\bf (of Theorem~\ref{thm:randbasis})}
Let $\rho$ be an irreducible $d_\rho$-dimensional
unitary representation of $G$.
We follow the notation of
Definition~\ref{def:rh1h2} for $\rho$.
Let $V''_2$ denote the orthogonal complement of $V'_1$ in 
$V'_1 + V'_2$. Let $\sigma_i$ denote the totally mixed state in
$V_i$ and $\sigma''_2$ denote the totally mixed state in
$V''_2 + (V_1 \cap V_2)$. By Proposition~\ref{prop:ovlap},
$\trnorm{\sigma_2 - \sigma''_2} < 
 2 \delta^{1/2} (1 - \delta^2)^{-1/4}$.
Let $\cB_\rho$ be a random orthonormal basis for $\rho$.
Let $P_i := P^{G, \cB_\rho}_{H_i}$ denote the probability 
distributions on the
vectors of $\cB_\rho$ got by Fourier sampling the states
$\ket{H_i}$ respectively, conditioned on $\rho$ 
being observed. Then $P_i$ is the probability distribution
got by measuring $\sigma_i$ in the basis $\cB_\rho$.
Let $P''_2$ denote the probability distribution
got by measuring $\sigma''_2$ in the basis $\cB_\rho$.
Then, 
$\totvar{P_2 - P''_2} < 
 2 \delta^{1/2} (1 - \delta^2)^{-1/4}$.
Define $r_{1,2} = \rank{\Pi_{1,2}}$. Note that
$r_i = r'_i + r_{1,2}$.

Suppose case~1 of Definition~\ref{def:rh1h2} applies. 
Let $W$ be a random orthonormal
set of $r'_1 + r'_2 + r_{1,2}$ vectors in $C^{d_\rho}$.
Define probability distributions $S$, $T$ with respect to $W$ as
in Lemma~\ref{lem:randspace1}. By symmetry, $P_1 = S$ and
$P''_2 = T$. Note that $r'_1 + r'_2 + r_{1,2} \leq r_1 + r_2$ and
$r'_1 + r'_2 + r_{1,2} \leq d_\rho < \sqrt{|G|}$. 
Hence,
$d_\rho \cdot \theta((r'_1 + r'_2 + r_{1,2})^{-3}) - 
 2 \log (r'_1 + r'_2 + r_{1,2}) = \Omega(\log^2 |G|)$.
The conditions of Lemma~\ref{lem:randspace1} are satisfied, and
we get, with probability at least $1 - \exp(-\Omega(\log^2 |G|))$
over the choice of $\cB_\rho$, that 
$\totvar{P_1 - P''_2} = \Omega\left(
 \frac{\sqrt{r'_1}}{r_1} + \frac{\sqrt{r'_2}}{r_2}
 \right)$.
Thus with probability at least $1 - \exp(-\Omega(\log^2 |G|))$
over the choice of $\cB_\rho$, 
\[
\totvar{P_1 - P_2} \geq
\totvar{P_1 - P''_2} - \totvar{P_2 - P''_2} \geq
\Omega\left(
\frac{\sqrt{r'_1}}{r_1} + \frac{\sqrt{r'_2}}{r_2}
\right) - 2 \delta^{1/2} (1 - \delta^2)^{-1/4}.
\]

Suppose case~2 of Definition~\ref{def:rh1h2} applies. 
Let $W$ be a random orthonormal
set of $r_1$ vectors in $C^{d_\rho}$.
Define probability distribution $S$ with respect to $W$ as
in Lemma~\ref{lem:randspace2}. By symmetry, $P_1 = S$.
Note that $r_1 \leq d_\rho < \sqrt{|G|}$. Hence,
$d_\rho \cdot \theta(r_1^{-2}) - 
 2 \log r_1 = \Omega(\log^2 |G|)$.
The conditions of Lemma~\ref{lem:randspace2} are satisfied, and
we get,
with probability at least $1 - \exp(-\Omega(\log^2 |G|))$
over the choice of $\cB_\rho$, 
that
$\totvar{P_1 - U} = \Omega\left(\frac{1}{\sqrt{r_1}}\right)$.
Also, $r_2 = \Omega(r_1 \log d_\rho)$. 
The conditions of Fact~\ref{fact:randspace} are satisfied, and
we get, 
with probability at least $1 - \exp(-\Omega(\log^2 |G|))$
over the choice of $\cB_\rho$, 
that
$\totvar{P_2 - U} = O\left(\frac{1}{\sqrt{r_1}}\right)$.
Thus 
with probability at least $1 - \exp(-\Omega(\log^2 |G|))$
over the choice of $\cB_\rho$, 
\[
\totvar{P_1 - P_2} \geq
\totvar{P_1 - U} - \totvar{P_2 - U} =
\Omega\left(\frac{1}{\sqrt{r_1}}\right).
\]

Suppose $|H_1| r_1 \geq |H_2| r_2$ i.\,e. $\hat{h} = |H_1| r_1$.
Then,
\[
\totvar{P^{G, \cB}_{H_1} - P^{G, \cB}_{H_2}} =
\sum_\rho \totvar{P^G_{H_1}(\rho) P^{G, \cB_\rho}_{H_1} -
                  P^G_{H_2}(\rho) P^{G, \cB_\rho}_{H_2}} =
\sum_\rho \frac{d_\rho}{|G|} \cdot 
          \totvar{H_1 r_1 P^{G, \cB_\rho}_{H_1} -
                  H_2 r_2 P^{G, \cB_\rho}_{H_2}}.
\]
Now,
\begin{eqnarray*}
\totvar{H_1 r_1 P^{G, \cB_\rho}_{H_1} -
        H_2 r_2 P^{G, \cB_\rho}_{H_2}}
&   =  & \totvar{H_1 r_1 (P^{G, \cB_\rho}_{H_1} - 
                          P^{G, \cB_\rho}_{H_2}) + 
                 (H_1 r_1 - H_2 r_2) P^{G, \cB_\rho}_{H_2}} \\
& \geq & \frac{H_1 r_1}{2} \cdot
         \totvar{P^{G, \cB_\rho}_{H_1} - P^{G, \cB_\rho}_{H_2}}. 
\end{eqnarray*}
The last step above follows from the facts 
$H_1 r_1 - H_2 r_2 \geq 0$, $\totvar{v} \geq \totvar{v_+}$ where
$(v_+)_i := v_i$ if $v_i \geq 0$, $(v_+)_i := 0$ otherwise, and
$\totvar{P_1 - P_2} = 2 \totvar{(P_1 - P_2)_+}$ for probability
vectors $P_1, P_2$. 
Also note that for any choice of representation bases $\cB$,
$\totvar{H_1 r_1 P^{G, \cB_\rho}_{H_1} -
         H_2 r_2 P^{G, \cB_\rho}_{H_2}} \geq |H_1 r_1 - H_2 r_2|$.
Hence, 
\begin{eqnarray*}
\totvar{P^{G, \cB}_{H_1} - P^{G, \cB}_{H_2}} 
& \geq & \sum_\rho \frac{d_\rho}{|G|} \cdot 
         \max\left\{
         \frac{H_1 r_1}{2} \cdot
         \totvar{P^{G, \cB_\rho}_{H_1} - P^{G, \cB_\rho}_{H_2}},
         |H_1 r_1 - H_2 r_2|
         \right\} \\
& \geq & \sum_\rho \frac{d_\rho}{|G|} r(G; H_1, H_2; \rho) \\
&   =  & r(G; H_1, H_2).
\end{eqnarray*}

For each representation $\rho$, the
confidence bound in applying the above random basis arguments 
is at least $1 - \exp(-\Omega(\log^2 |G|))$. Since there are
at most $|G|$ representations, the total confidence bound is
at least $1 - |G| \exp(-\Omega(\log^2 |G|)) =
1 - \exp(-\Omega(\log^2 |G|))$. 
This completes the proof of the theorem.
\end{proof}

We now have all the tools to prove that $r(\cHp) = \Omega(1)$.
In fact, we can now prove the following theorem. 
\begin{theorem}
\label{thm:heisenberg}
The random strong method is sufficient to solve the hidden subgroup
problem in the Heisenberg group $\cHp$. The query complexity of
this algorithm is $O(\log p)$. The quantum part of the algorithm
consists of a circuit of size $O(\log^4 p)$ followed by a circuit
of size $\tilde{O}(p^2)$ for implementing the measurement in a random
orthonormal basis. The classical
post-processing does not make any queries and has a running time of 
$\tilde{O}(p^4)$.
\end{theorem}
\begin{proof}
First, we characterize the normal core families in the
Heisenberg group. We have that 
\[
\corefamily{\cHp} = \{{\cal H}_p\}, ~~~
\corefamily{\zeta(\cHp)} = \{\zeta(\cHp)\}, ~~~
\corefamily{N_i} = \{N_i\}, \,\mbox{for } 
i \in \{0, \ldots, p-1, \infty\}
\]
are families of size $1$ each. For the trivial group we get that
\[
\corefamily{\{1\}}=\{A_{i,j}: i \in \{0,\ldots,p-1,\infty\}, 
                              j \in \{0, \ldots, p-1\}\} \cup
                   \{\{1\}\}. 
\]
If $H_1$ and $H_2$ are candidate hidden subgroups
from different normal core families, then by 
Proposition~\ref{prop:normcore} we get that 
$r(\cHp; H_1, H_2) \geq w(\cHp; H_1, H_2) \geq 1/2$. 
We now consider the situation where both
$H_1, H_2 \in \corefamily{\{1\}}$. 
We fix an irreducible representation
$\rho = \rho_k$ (for $k=1, \ldots, p-1$) of degree $\deg \rho =
p$. Now, we distinguish two cases: 
\begin{enumerate}
\item $|H_1|=|H_2|=p$, i.\,e., there
      are $i$, $j$, $i'$, $j'$, $(i, j) \neq (i', j')$
      such that $H_1 = A_{i,j}$ and
      $H_2 = A_{i',j'}$. Using the notation of 
      Definition~\ref{def:rh1h2} we have that $r_1=r_2=1$, 
      since by Lemma~\ref{lem:ovlapHeisen} the ranks of 
      $P_{k;i,j}$ and $P_{k; i',j'}$ are both one. 
      Also, $\norm{P_{k;i,j} P_{k;i',j'}} \leq \frac{1}{\sqrt{p}}$.
      This also implies that $r'_1 = r_1 = 1$ and $r'_2 = r_2 = 1$,
      since the one-dimensional
      projectors $P_{k;i,j}$, $P_{k;i',j'}$ are linearly independent.
      Hence, $P'_{k;i,j} = P_{k;i,j}$ and 
      $P'_{k;i',j'} = P_{k;i',j'}$.
      So, $\delta=\norm{P'_{k;i,j} P'_{k;i',j'}} \leq 
                  \frac{1}{\sqrt{p}}$.
      Since $|H_1| r_1 = |H_2| r_2 = p$, $\hat{h} = p$ and 
      $\tilde{h} = 0$. As 
      \[
      \frac{{\sqrt{d_\rho}}}{{\log |\cHp|}}
      = \frac{\sqrt{p}}{3 \log p} 
      = \Omega((r_1+r_2)^{3/2}) 
      = \Omega(1),
      \]
      we are in the first case of Definition~\ref{def:rh1h2} 
      and obtain
      that 
      \[
      r(\cHp; H_1, H_2; \rho) = \frac{p}{2} \cdot
      \left(\Omega(1) - 2 (p - 1)^{-1/4}\right) = 
      \Omega(p). 
      \]

\item $|H_1|=p$ and $|H_2|=1$, 
      i.\,e., we have to
      distinguish $H_1 = A_{i,j}$ from the trivial subgroup 
      $H_2=\{1\}$. 
      In this case $r_1=1$ and $r_2 = \rank{\rho(\{1\})} = p$ which 
      implies that $\hat{h} = p$, $\tilde{h} = 0$. Since
      \[
      \frac{\sqrt{d_\rho}}{\log |\cHp|} = \Omega(r_1) 
      ~~~ \mbox{and} ~~~ 
      \frac{r_2}{r_1} = p = \Omega(\log^2 |\cHp|), 
      \]
      we are in the second case of Definition~\ref{def:rh1h2} 
      and obtain that 
      \[
      r(\cHp; H_1, H_2; \rho) =
      \frac{p}{2} \cdot \Omega(1) = 
      \Omega(p).
      \]
\end{enumerate}
Overall we obtain that for $H_1, H_2 \in \corefamily{\{1\}}$,
\[
r(\cHp; H_1, H_2) = 
\sum_\rho \frac{d_\rho}{p^3} \cdot r(\cHp; H_1, H_2; \rho) \geq 
\sum_{k=1}^{p-1} \frac{p}{p^3} \cdot r(\cHp; H_1, H_2; \rho_k)\geq 
\frac{(p - 1) p}{p^3} \cdot \Omega(p) = 
\Omega(1). 
\]
Hence, $r(\cHp) = \Omega(1)$. Recall that $s(\cHp) = O(p^2)$.
Now Corollary \ref{cor:randbasis} shows that with
probability at least $2/3$ over the choice of
random representation bases,
the HSP for $\cHp$ can be solved using $O(\log p)$
random strong Fourier samplings.

As shown in Proposition~\ref{prop:qftcHp}, the $\QFT$ over $\cHp$ can
be implemented using $O(\log^3 p)$ elementary quantum gates. Since
there are $O(\log p)$ Fourier samplings, the initial part of the
quantum circuit has size $O(\log^4 p)$.
The claimed statements about the number of quantum operations
necessary to implement a measurement in a random orthonormal basis 
follow from general upper bounds of $\tilde{O}(p^2)$ on 
the number of gates in a
factorization of a unitary operation $U \in \U(p)$ into elementary
gates. The classical time to generate the random $U$ is
$\tilde{O}(p^3)$ since we can start with a set of $p$ random unit
vectors and apply Gram-Schmidt orthonormalisation to obtain a random
unitary matrix. For the classical post-processing we have to compute a
table of probability distributions with respect to the random
measurement bases for all subgroups.  Since there are $O(p^2)$
subgroups and each probability distribution computation takes time
$\tilde{O}(p^2)$ we can upper bound this by $\tilde{O}(p^4)$. After
this table has been precomputed the actual algorithm to find the
hidden subgroup is `minimum-finding-like' in which we `compare' two
subgroups at a time. This takes time $\tilde{O}(p^2)$.  Overall, we
obtain that the running time of the classical part of this algorithm
can be upper bounded by $\tilde{O}(p^4)$.
\end{proof} 

%%%%%%%%%%%%%%%%%%%%%%%%%%%%%%%%%%%%%%%%%%%%%%%%%%%%%%%%%%%%
%
% Section: Conclusions
%
%%%%%%%%%%%%%%%%%%%%%%%%%%%%%%%%%%%%%%%%%%%%%%%%%%%%%%%%%%%%

%\section{Conclusions}
%\cite{EWSLC:2003}

%%%%%%%%%%%%%%%%%%%%%%%%%%%%%%%%%%%%%%%%%%%%%%%%%%%%%%%%%%%%
%
% Acknowledgements
%
%%%%%%%%%%%%%%%%%%%%%%%%%%%%%%%%%%%%%%%%%%%%%%%%%%%%%%%%%%%%

\subsection*{Acknowledgments}
We thank Fr\'{e}d\'{e}ric Magniez, Leonard Schulman, Cris Moore,
Alex Russell and Avery Miller for useful discussions.

%%%%%%%%%%%%%%%%%%%%%%%%%%%%%%%%%%%%%%%%%%%%%%%%%%%%%%%%%%%%
%
% The Literature
%
%%%%%%%%%%%%%%%%%%%%%%%%%%%%%%%%%%%%%%%%%%%%%%%%%%%%%%%%%%%%

\bibliographystyle{alpha} 
\newcommand{\etalchar}[1]{$^{#1}$}

\end{document}